\documentclass[onecolumn]{emulateapj}

\newcommand{\myemail}{gokhale@baton.phys.lsu.edu}
\renewcommand{\d}{{\rm d}}
\newcommand{\sgn}{{\rm sgn}}

\shorttitle{Evolution of Double White Dwarfs}
\shortauthors{Gokhale et al.}

\begin{document}
\title{Evolution of Close White Dwarf Binaries}
\author{Vayujeet Gokhale, Xiao Meng Peng, Juhan Frank}
\affil{Department of Physics and Astronomy,\\ Louisiana State
University, Baton Rouge, LA 70803-4001.\\ \myemail.}

\begin{abstract}
We describe the evolution of double degenerate binary systems,
consisting of components obeying the zero temperature mass radius
relationship for white dwarf stars, from the onset of mass transfer
to one of several possible outcomes including merger, tidal
disruption of the donor, or survival as a semi-detached AM CVn
system. We use a combination of analytic solutions and numerical
integrations of the standard orbit-averaged first-order evolution
equations, including direct impact accretion and the evolution of
the components due to mass exchange. We include also the effects of
mass-loss during super-critical (super-Eddington) mass transfer and
the tidal and advective exchanges of angular momentum between the
binary components. We cover much the same ground as \citet{Maet04}
with the additional effects of the advective or consequential
angular momentum from the donor and its tidal coupling to the orbit
which is expected to be stronger than that of the accretor. With the
caveat that our formalism does not include an explicit treatment of
common envelope phases, our results suggest that a larger fraction
of detached double white dwarfs than what has been hitherto assumed,
survive the onset of mass transfer, even if this mass transfer is
initially unstable and rises to super-Eddington levels. In addition,
as a consequence of the tidal coupling, systems that come into
contact near the mass transfer instability boundary undergo a phase
of oscillation cycles in their orbital period (and other system
parameters). Unless the donor star has a finite entropy such that
the effective mass-radius relationship deviates significantly from
that of a zero temperature white dwarf, we expect our results to be
valid. Much of the formalism developed here would also apply to
other mass-transferring binaries, and in particular to cataclysmic
variables and Algol systems.
\end{abstract}


\section{Introduction}
\label{Intro} Binary white dwarfs which are close enough to be
driven together by angular momentum losses due to gravitational
radiation, and perhaps additional mechanisms, will undergo a phase
of mass transfer during which the ultimate fate of the binary is
decided. There are many reasons to revisit these dynamical phases of
the evolution of double degenerate binaries today. Chief among these
is that a survey of the published literature on the subject reveals
that many pieces of the puzzle have been explored in different
degrees but we still lack a uniform theoretical understanding of the
fate of these binaries for all possible mass ratios, orbital
parameters and origins. Ideally, one would like our theoretical
understanding to be such that given a white dwarf binary of
arbitrary masses and compositions at the time that the less massive
component gets into contact, one could reconstruct the previous
evolutionary pathways and the subsequent evolution to merger, tidal
disruption or stable mass transfer.

The reason for this lack of uniformity is that understanding
double white dwarf binaries (DWD) in this rapid phase of
interaction can be quite complex and demanding, necessitating
detailed hydrodynamics, nuclear physics, radiative transfer and
stellar structure. Therefore it is natural that different
assumptions and techniques are used when one attempts to answer a
question relevant to different classes of phenomena. For example,
the approaches adopted in studying the putative progenitors of
supernovae of type Ia, or the sources of gravitational waves for
LISA, or the progenitors of AM CVn binaries, are all very
different. While we shall not attempt to cover all the rich
physics that may be ultimately necessary to have a uniform and
reliable treatment of all possible outcomes of the interaction, we
will describe these interactions within a single semi-analytic
framework, making connections where appropriate to well-known
results already in the literature.

 Our particular motive for having developed the
understanding described in this paper is that our group has been
improving and running a 3-D numerical hydrocode \citep{MTF, DMTF}
which is currently capable of following self-consistently the
evolution of model white dwarf binaries through these rapid phases
of mass transfer, tracking mass and angular momentum with high
accuracy for over 30 orbital periods. In the course of numerous
evolutions in which we controlled the rate of driving by angular
momentum losses, or prescribed an arbitrary rate of pseudo-thermal
expansion of the donor, it became clear that we needed some simpler
insight into the behavior of the models. This led us to extend the
analytic solution of \citet{WebIb} (WI) for the time-dependent
behavior of the mass transfer by relaxing most of the assumptions
made to render the problem tractable. Here, we retain the
simplifying assumption of Roche geometry, but allow all the binary
parameters to vary self-consistently. Thus we investigate
numerically a system of first-order evolution equations for a
variety of cases and recover the WI solution when appropriate
conditions are applied. We then extend our insight to more general
cases, and comment on the applicability and limitations of our
approach. While we are still far from the more ambitious goal
described above, we think that the present investigation may also
provide a useful framework for other workers in the field of binary
evolution and especially for those interested in large-scale
numerical simulations of these binaries. Many of the techniques and
theoretical insights in this paper are applicable to other
semi-detached binaries and may have some relevance to contact
binaries.

\section{Basic Equations}
\label{Eqns} Consider a binary having nearly spherical components of
masses $M_1$ and $M_2$ and separation $a$. Without loss of
generality, we shall assume that as this binary evolves from a
detached to a semi-detached state, star 2 is the one that fills its
critical Roche lobe and becomes the donor. We define the mass ratio
of the binary to be $q=M_2/M_1$.  Assuming for simplicity that the
spin axes of the individual components of the binary are
perpendicular to the orbital plane, and that the orbit is circular,
the total angular momentum of the system is given by:
\begin{eqnarray}
\nonumber J_{\rm tot} &=& J_{\rm orb} + J_1 + J_2 \\
                  &=& M_1 M_2 \left(G a/M \right)^{1/2} +
                  k_1 M_1 R_1^2~ \omega_1 + k_2 M_2 R_2^2
                  ~\omega_2 ,
                  \label{Jtot}
\end{eqnarray}
where $k_i$ are dimensionless constants depending on the internal
structure of the components and $\omega_i$ are the angular spin
frequencies. The first term in eq. (\ref{Jtot}) represents the
orbital angular momentum of the components, and the two other terms
represent the spin angular momenta of the stars. The form of the
orbital angular momentum term adopted above assumes the binary
revolves at the Keplerian frequency $\Omega = (GM/a^3)^{1/2}$, which
is a good approximation if the stars are centrally condensed.

For a fully synchronous configuration, it is easy to show that the
total angular momentum and the total energy have a minimum at the
same separation $a_{\rm min}= [3(I_1+I_2)/\mu]^{1/2}$, where the
$I_i = k_i M_i R_i^2$ are the moments of inertia of the components,
and $\mu$ is the reduced mass. Even if the orbital frequency and the
spin frequency are not synchronized but remain proportional to one
another ($\omega_i=f_i \Omega$), there is a minimum of $J_{\rm tot}$
at some $a_{\rm min}$. Also the total energy of the system will have
a minimum but in general it will not occur at $a_{\rm
min}$.\footnote{LRS2 show that for Riemann-S and Roche-Riemann
sequences $\d E = \Omega\d J + \Lambda\d{\cal C}$, where $\Lambda$
is the angular velocity of internal motions and ${\cal C}$ is the
equatorial circulation. Thus, as a binary evolves driven by
gravitational wave radiation, circulation is conserved and the
minima of $E$ and $J$ will coincide. However tidal dissipation does
not preserve ${\cal C}$ and thus in general the minima will {\em
not} coincide in the presence of tidal spin-orbit coupling.}

 For our purposes it will be sufficient to work with the
approximate form of eq. (\ref{Jtot}) given above. A more complete
and thorough discussion of the secular and dynamical stability of
polytropic binaries has been presented in two well-known series of
papers by Lai, Rasio \& Shapiro (LRS1-LRS5) and further developed
with SPH simulations in papers by Rasio \& Shapiro (RS1-RS3), who
also addressed the role of mass-transfer.

If the donor has a relatively soft equation of state and if $q\ne
1$, the binary tends to become semi-detached and mass-transfer
occurs before it falls prey to the tidal instability mentioned
above. Mass transfer changes the initial configuration as the system
evolves and may either drive the system to smaller separations and
thus closer to the onset of the tidal instability or to larger
separations and toward stability. Similarly, mass loss from the
system can affect the dynamic stability of the system. We
investigate the behavior of such systems (in particular DWD
binaries) in the cases of conservative and non-conservative mass
transfer, driven by gravitational wave radiation (GWR).

\subsection{Orbital and spin angular momentum}
\label{AML}

A system of two point masses orbiting around each other, in circular
orbits, radiates gravitationally \citep{LaLi}. The loss of orbital
angular momentum as a result is given by
\begin{equation}
\Big(\frac{\dot J}{J}\Big)_{\rm GWR} = ~ -\frac{32}{5}
\frac{G^3}{c^5} \frac{M_1 M_2 M}{a^4} \label{JdotGWR}
\end{equation}
Although GWR is likely to be the only important mode of angular
momentum loss from DWD binaries, for any general form of systemic
angular momentum loss $\dot J_{\rm sys}$, we can re-arrange eq.
(\ref{Jtot}) to obtain
\begin{eqnarray}
J_{\rm orb} = J_{\rm tot} -(J_1+J_2) \Rightarrow \dot J_{\rm orb} =
\dot J_{\rm sys} -(\dot J_1 + \dot J_2)\, , \label{Jorb}
\end{eqnarray}
allowing for the possibility of spin-orbit coupling of the angular
momenta. The rate of change of the spin angular momentum of each
individual star can be given as the sum of the advected or {\it
consequential} angular momentum transport plus the effect of the
tidal torques
\begin{eqnarray}
\dot J_1 = \dot M_1 j_{1} + \dot J_{\rm 1, tid}\ , \label{J1dot}
\\
\dot J_2 = \dot M_2 j_{2} + \dot J_{\rm 2, tid}\ . \label{J2dot}
\end{eqnarray}
The term denoting the advected component from the donor has not been
included in previous treatments concerning white dwarf donors
\citep{Maet04}, but has been discussed in the case of evolved donors
(see for example, \cite{PratStritt76} and \cite{Sav78}). For this
reason, a few remarks clarifying the meaning of the consequential
terms are in order. In the above equations $j_1$ and $j_2$ indicate
the specific angular momenta of the matter {\em arriving} at the
accretor and the matter {\em leaving} the donor respectively. In a
conservative system these will refer to the specific angular
momentum of the {\em same} material with respect to the center of
mass of each star, but at {\em different} times. We assume that as
the stream leaves the donor there is no back torque that could
modify the spin of the donor. Therefore $j_2$ is entirely determined
by the instantaneous conditions at the donor. However, the material
traveling in the stream experiences a time varying torque due to the
binary that changes $j_1$ at the expense of the orbital angular
momentum alone i.e. not torquing the spins of either component.
These are the assumptions underlying the standard calculations of
$j_1$ and the estimates of the circularization radius which go back
to \citet{Fla75} and \citet{LuSh75}. We will discuss later what are
the appropriate values for $j_1$ and $j_2$, and write down just the
general expressions here. Substituting eqs. (\ref{J1dot}) and
(\ref{J2dot}) in eq. (\ref{Jorb}) and rearranging
\begin{eqnarray}
\dot J_{\rm orb} = \dot J_{\rm sys} - \left[- \dot M_2(j_1-j_2) +
\dot J_{\rm 1, tid}+ \dot J_{\rm 2, tid}\right]\ , \label{JorbDot}
\end{eqnarray}
where we have assumed conservative mass transfer.
Also, we can write the tidal torque as
\begin{eqnarray}
\dot J_{\rm 1, tid} = \frac{k_1 M_1 R_1^2}{\tau_{\rm s_1}} (\Omega -
\omega_1) \label{J1dottid}
\\
\dot J_{\rm 2, tid} = \frac{k_2 M_2 R_2^2}{\tau_{\rm s_2}} (\Omega -
\omega_2) \label{J2dottid}
\end{eqnarray}
where $\tau_{\rm s_1}$ and $\tau_{\rm s_2}$ are the synchronization
timescales of the accretor and donor respectively (See Section
\ref{cycles}). Eq. (\ref{JorbDot}) becomes
\begin{eqnarray}
\dot J_{\rm orb} = \dot J_{\rm sys} +\dot M_2 (j_1-j_2) - \frac{k_1
M_1 R_1^2}{\tau_{\rm s_1}} (\Omega - \omega_1) - \frac{k_2 M_2
R_2^2}{\tau_{\rm s_2}} (\Omega - \omega_2) \label{JorbDot a}
\\
= \dot J_{\rm sys} + \frac{\dot M_2}{M_2}\Big[M_2(j_1-j_2) + k_1 M_1
R_1^2 \frac{\tau_{{M_2}}}{\tau_{\rm s_1}} (\Omega - \omega_1) + k_2
M_2 R_2^2 \frac{\tau_{{M_2}}}{\tau_{\rm s_2}}(\Omega-\omega_2)\Big]
\label{JorbDot b}
\end{eqnarray}
where $\tau_{M_2} = - M_2/\dot{M_2}$ is the mass transfer time
scale. In eq. (\ref{JorbDot b}) the tidal torques have been placed
inside the brackets although, strictly speaking, they are not
consequential. In most cases encountered in cataclysmic variables
(CVs) and low-mass X-ray binaries (LMXBs), $\tau_{\rm
s_1}\gg\tau_{M_2}$ and $\tau_{\rm s_2}\ll\tau_{M_2}$, and thus
usually $\omega_1\gg\Omega$, while $\vert (\Omega - \omega_2)/\Omega
\vert \sim \tau_{\rm s_2}/\tau_{M_2}$. However, in double degenerate
binaries there is considerable uncertainty about the synchronization
timescales, and in numerical simulations of mass transfer
$\tau_{M_2}$ is orders of magnitude shorter than the typical
timescales one encounters in long-term accreting binaries. Therefore
we will retain both tidal terms and investigate their effect on
dynamical evolutions. Note that our eq. (\ref{JorbDot a}) is
equivalent to eq. (1) of \citet{Maet04}, with two extra terms
arising from the advective and tidal contributions from the donor.

\subsection{Binary separation}
\label{binsep}
We now derive the equations for the evolution of the
binary separation, the radius of the donor  and the Roche lobe
radius using the above equations. From the functional form of the
orbital angular momentum, we know that for a conservative system
\begin{eqnarray*}
\Big(\frac{\dot J}{J}\Big)_{\rm orb} = \frac{\dot M_2}{M_2}(1-q)
+\frac{1}{2} \frac{\dot a}{a} .
\end{eqnarray*}
Comparing this with eq. (\ref{JorbDot a}) and rearranging, we obtain
\begin{eqnarray}
\frac{\dot a}{2a} = \frac{\dot J_{\rm sys}}{J_{\rm orb}} - \frac{
\dot M_2}{M_2} \Big[1 - q - M_2\frac{j_1-j_2}{J_{\rm orb}}\Big] -
\frac{k_1 M_1 R_1^2}{J_{\rm orb}\tau_{\rm s_1}} (\Omega - \omega_1)
- \frac{k_2 M_2 R_2^2}{J_{\rm orb}\tau_{\rm s_2}} (\Omega-\omega_2)
\label{adot}
\end{eqnarray}
The sign of the quantity in brackets determines whether mass
transfer tends to expand the system and thus oppose the effect of
angular momentum losses, or lead to enhanced contraction. The change
of sign will generally occur at some mass ratio $q_a$ whose value we
discuss below. Symbolically we may write
\begin{eqnarray}
\frac{\dot a}{2a} &=& \frac{\dot J_{\rm sys}}{J_{\rm orb}}
-\frac{\dot J_{\rm 1, tid}+ \dot J_{\rm 2, tid}}{J_{\rm orb}} -
\frac{\dot M_2}{M_2} [q_a - q] \label{qa}
\\
q_a &\equiv& 1- M_2\frac{j_1-j_2}{J_{\rm orb}} , \label{qadef}
\end{eqnarray}
Thus (remembering that $\dot M_2$ is intrinsically negative) if
$q>q_a$, mass transfer will contribute to reducing the separation
and, as we will see below, tend to make the binary more unstable to
mass transfer. On the other hand, if $q<q_a$, mass transfer will
oppose the effects of driving and will tend to stabilize the binary.
As $q$ decreases on mass transfer, it is possible that a system
which started life with $q>q_a$ may evolve to a more stable
configuration, if it does not fall prey to the tidal instability. In
eq. (\ref{qa}) the tidal synchronization torques may be considered
additional contributions to the driving and they may subtract or add
angular momentum to the orbit depending on the case. Note that $q_a$
should be interpreted as the mass ratio at which the last term in
eq. (\ref{qa}) changes sign. Its value can be estimated from the
initial mass ratio since it is a slowly varying function of $q$ in
the conservative case, but in general it is obtained
self-consistently as the binary evolution is followed numerically.

The second term in the definition of $q_a$ represents the effects of
the net consequential transfer of angular momentum from orbit to
spins. We introduce the symbol $\zeta_{\rm c} = M_2(j_1-j_2)/J_{\rm
orb}$ for this term, noting that it stands for the consequential
contribution to  $-\d\log{J_{\rm orb}}/\d\log{M_2}$. For direct
impact accretion $j_1=j_{\rm circ} (\approx b_1^2 \Omega)$ is the
specific angular momentum carried by the stream as it hits the
accretor. The approximate value in parenthesis is valid for a
synchronous donor and $b_1=a(0.5-0.227\log{q})$ is the distance from
the center of mass of the accretor to the inner Lagrangian point
$L_1$ \citep{FKR}. With the standard definition of circularization
radius $r_h= R_{\rm circ}/a$ \citep{Fla75, LuSh75, VeRa, Maet04},
$M_2 j_{\rm circ}/J_{\rm orb} = [(1+q)r_h]^{1/2}$. Note, however,
that the tidal coupling of the donor spin to the orbit was neglected
by \citet{Maet04}, and the consequential term proportional to
$j_2\approx R_2^2 \omega_2$ was not included either. The precise
value of $j_2$ depends on the details of the flow in the vicinity of
$L_1$ in a non-synchronous donor (See \cite{Krus63} and
\cite{CsatSko05}). For the purposes of this investigation we will
simply adopt $j_2=R_2^2 \omega_2$. With these definitions we may
write
\begin{equation}
q_a = 1 - \zeta_{\rm c} = 1
-[(1+q)r_h]^{1/2}(1-\frac{R_2^2\omega_2}{\sqrt{G M_1 R_{\rm circ}}}) .
\label{explicitqa}
\end{equation}
The net effect of the consequential redistribution of angular
momentum in the binary depends on the sign of $\zeta_{\rm c}$. For
DWD binaries, $j_1>j_2$ during the direct impact stage, and this holds
even after the onset of disk accretion, when $j_1=\sqrt{G M_1 R_1}$ and
$\zeta_{\rm c}$ becomes smaller but remains positive.
In cataclysmic variables and low-mass X-ray binaries, the accretion
disk returns via tides most of the angular momentum advected by the
stream, the donor is almost synchronous, and the tidal coupling of
the accretor to the orbit is very weak. However, in this case $R_1\ll a$ and thus it is
more likely that $j_2>j_1$. While all the additional terms in
eqs. (\ref{qa}, \ref{qadef}) are relatively small, yielding $q_a\approx 1$,
in some cases $\zeta_{\rm c}<0$, and thus
$q_a\ga 1$ making these systems
slightly more stable.

The expression (\ref{explicitqa}) obtained above proves very useful
in the interpretation of results of large-scale numerical
hydrodynamic simulations of the dynamical evolution of binaries
undergoing mass transfer, with and without driving by angular
momentum losses \citep{DMTF}(see also \S\ref{hydrocomp}).

The Roche lobe radius for the donor is accurately given by the
formula due to \citet{Egg83}
\begin{eqnarray}
r_{\rm L}\equiv\frac{R_{\rm L}}{a} = \frac{0.49 q^{2/3}}{0.6 q^{2/3}
+ \ln (1+q^{1/3})} \label{Egg form}
\end{eqnarray}
and so with the notation of \citet{Maet04}
\begin{eqnarray*}
\nonumber \frac{\dot R_{\rm L}}{R_{\rm L}} = \zeta_{r_{\rm L}}
\frac{\dot M_2}{M_2} + \frac{\dot a}{a}\, ,
\end{eqnarray*}
where $\zeta_{r_{\rm L}}\approx 1/3$ is the logarithmic derivative
of $r_{\rm L}$ with respect to $M_2$\footnote{In the range $0<q\le
1$, the function $\zeta_{r_{\rm L}}$ takes values between 0.32 and
0.46, and is well approximated by $\zeta_{r_{\rm L}}\approx
0.30+0.16 q$ for $0.1\le q\le 1$}. Collecting results, we get
\begin{eqnarray}
\frac{\dot R_{\rm L}}{R_{\rm L}}= \frac{2 \dot J_{\rm sys}}{J_{\rm
orb}}-2\frac{\dot J_{\rm 1, tid}+ \dot J_{\rm 2, tid}}{J_{\rm orb}}
-\frac{2 \dot M_2}{M_2}[q_a-\frac{\zeta_{r_{\rm L}}}{2} -q]
\label{RLdot}
\end{eqnarray}
Generalizing the meaning of the symbols introduced by \citet{WebIb}
to include tidal and consequential terms, we can write the
equivalent expressions
\begin{eqnarray} \frac{\dot R_{\rm L}}{R_{\rm
L}} &=& \nu_{\rm
L}+\zeta_{\rm L}\frac{\dot M_2}{M_2}\\
\nu_{\rm L} &=& \frac{2 \dot J_{\rm sys}}{J_{\rm orb}}-2\frac{\dot
J_{\rm 1, tid}+ \dot J_{\rm 2, tid}}{J_{\rm orb}}\\
\zeta_{\rm L} &=& -2 q_a+\zeta_{r_{\rm L}}+2q, \label{RLdotarr}
\end{eqnarray}
where the symbol $\nu$ stands for driving terms and the $\zeta$ for
logarithmic derivatives with respect to donor mass. In the same
spirit we write the logarithmic time derivative of the donor radius
$R_2 \equiv R_2 (M_2,t)$ as:
\begin{equation}
\frac{\dot R_2}{R_2} = \nu_2 +\zeta_2 \frac{\dot M_2}{M_2}
\label{R2dot}
\end{equation}
where $\nu_2$ = $(\partial \ln R_2/\partial t)_{M_2}$ represents the
rate of change of the donor radius due to intrinsic processes such
as thermal relaxation and nuclear evolution, whereas $\zeta_2$
usually describes changes resulting from adiabatic variations of
$M_2$ \citep{DMR}. More generally, since the radial variations due
to the above mentioned effects operate on different timescales, it
is more appropriate to think of $\zeta_2$ as the effective
mass-radius exponent, averaged over the characteristic timescale of
mass transfer. If the donors are degenerate as in the WD case, or if
thermal relaxation is sufficiently rapid, $\zeta_2$ is simply
obtained from the equilibrium mass-radius relationship for the
donor. But in non-degenerate donors, if the thermal relaxation
timescale $\tau_{\rm th}$ becomes comparable to the mass transfer
timescale $\tau_{M_2}= M_2/\dot M_2$, as is thought to occur during
the evolution of CVs, the effects of thermal lag in the donor radius
become important, and $\zeta_2$ deviates from the equilibrium value
(See Appendix \ref{Appzeta}).

\section{Mass transfer rate} \label{masstran}

The following discussion of mass transfer and its stability is rooted in
a similar treatment of mass transfer under consequential angular momentum
losses \citep{KiKo95}.
For all types of donor star, the mass transfer rate is a strong
function of the depth of contact, defined here as the amount by
which the donor overflows its Roche lobe $\Delta R_2\equiv
R_2-R_{\rm L}$, suitably normalized. We adopt the following
expression valid for most cases of interest:
\begin{equation}
\dot M_2 = - \dot M_0(M_1, M_2, a) f(\Delta R_2), \label{M2dot}
\end{equation}
where $\dot M_0$ is a relatively gentle function of binary
parameters, while $f$ is a rapidly varying dimensionless function of
$\Delta R_2$. For example, for polytropic donors with index $n$, $f=
(\Delta R_2/R_2)^{n+3/2}$ \citep{PaSi}; while for a donor with an
atmospheric pressure scale height $H$, $f=\exp{(\Delta R_2/H)}$ is
appropriate \citep{Ri88}. In fact, the exact value of the
normalization rate $\dot M_0(M_1, M_2, a)$ is not important at all
in steady state because the equilibrium rate is determined by the
rate of driving: given a particular value of $\dot M_0$, the depth
of contact will adjust to yield the transfer rate sustainable by the
driving. In transient situations, if the mass transfer is varying
rapidly, or if the depth of contact becomes large, the normalization
becomes relevant.

In principle, Eqs. (\ref{J1dottid}), (\ref{J2dottid}), (\ref{qa})
and (\ref{RLdot})-(\ref{M2dot}) completely specify the system and
can be numerically integrated, and we discuss some examples of such
evolutions later in \S\ref{evol}. However, before proceeding, it is
more illuminating to analyze the general implications of eq.
(\ref{M2dot}). Under the assumptions mentioned above
\begin{equation}
\ddot M_2 = -\dot M_0\frac{\partial f}{\partial\Delta
R_2}\Big(\frac{\dot R_2}{R_2}-\frac{\dot R_{\rm L}}{R_{\rm L}}\Big)
, \label{M2dbldot}
\end{equation}
where we have neglected the slower variations of $\dot M_0$ with
system parameters and we have also assumed that the depth of contact
is small $\Delta R_2<<R_2$. Thus the mass transfer rate will be
steady whenever the size of the donor varies in step with its Roche
lobe. From Eqs. (17)-(20) and (\ref{M2dbldot}) we obtain the
equilibrium mass transfer rate
\begin{eqnarray}
\Big(\frac{\dot M_2}{M_2}\Big)_{\rm eq} = \frac{
\nu_{\rm L}-\nu_2} {2(q_{\rm stable}-q)}= \frac{\nu_{\rm L} -
\nu_2}{\zeta_2-\zeta_{\rm L}} \label{M2doteq}
\end{eqnarray}
where
\begin{equation}
q_{\rm stable} = q_a -\frac{\zeta_{r_{\rm L}}}{2} +\frac{\zeta_2}{2}
, \label{qstab}
\end{equation}
is the critical mass ratio for stability of mass transfer. The
alternative expression on the r.h.s of eq. (\ref{M2doteq}) thus has
the same form as in \citet{WebIb}, except that here the driving and
consequential terms include the effects of tidal coupling and direct
impact accretion. Furthermore, we will allow these terms to vary
self-consistently as the evolution proceeds (See \S\ref{evol}).

For example, if we assume that the binary is synchronized, the orbit
circular, the donor is a polytrope of index $n=3/2$, and $\nu_2=0$
(no thermal or nuclear evolution), then $\zeta_2 \sim -1/3$,
$\zeta_{r_{\rm L}} \sim 1/3$, and  we get
\begin{eqnarray*}
\Big(\frac{\dot M_2}{M_2}\Big)_{\rm eq} = \frac{\dot J_{\rm
sys}/J_{\rm orb}}{2/3 -\zeta_{\rm c}- q}
\end{eqnarray*}
which is the familiar form in the case of direct impact DWD's
\citep{Maet04}, except that here $\zeta_{\rm c}$ is reduced by the
contribution from the donor as given by eq. (\ref{explicitqa}).

The equilibrium mass transfer rate was obtained by demanding that
$\ddot M_2=0$ in eq. (\ref{M2dbldot}).  In general, if $\dot
M_2\neq(\dot M_2)_{\rm eq}$, one can rewrite this equation as
follows:
\begin{equation}
\ddot M_2 = -2\frac{\dot M_0}{M_2} \frac{\partial f}{\partial\Delta
R_2} (q_{\rm stable}-q)\left[\dot M_2 - (\dot M_2)_{\rm eq}\right].
\label{stab}
\end{equation}
The sign of the pre-factor on the r.h.s. is negative if $q<q_{\rm
stable}$, thus $\dot M_2$ will tend toward the equilibrium value and
mass transfer is stable. If $q>q_{\rm stable}$ no attainable
equilibrium mass transfer exists and mass transfer is unstable.

\section{Analytic Solutions}
\label{analytic} Assuming that most of the parameters characterizing
the binary and the donor remain constant during (the usually much
faster) evolution of the accretion rate, it is possible to obtain
analytic solutions for the evolution of the accretion rate itself.
In this section we generalize the result of \citet{WebIb} to include
donors with an arbitrary polytropic index and with an isothermal
atmosphere. These can be later compared to our integrations of the
evolution equations in which we allow all binary and donor
parameters to vary self-consistently and also with the results of
large-scale hydrodynamic simulations \citep{DMTF}.

Assuming that the donor in the binary can be represented by a
polytrope, the mass transfer rate is given by a formula derived by
\citet{Jedr} assuming laminar flow, and quoted by \citet{PaSi}
\begin{equation}
-\dot M_2 = \dot M_0 \Big(\frac{R_2-R_{\rm L}}{R_2}\Big)^{n+3/2}
\end{equation}
Raising both sides of the above equation to the power $2/(2n+3)$ and
differentiating, we obtain
\begin{equation}
\frac{d}{dt}(-\dot M_2)^{2/2n+3} = (\dot M_0)^{2/2n+3} [(\nu_2 -
\nu_{\rm L})+ \frac{\dot M_2}{M_2}(\zeta_2-\zeta_{\rm L})],
\label{dotM2power}
\end{equation}
where we have set the factor $R_{\rm L}/R_2$ to unity, given that in
most situations $\Delta R_2 \ll R_2$. The analytic solutions
discussed here assume that the driving rate $\nu_{\rm L}$, the
intrinsic radial variation rate $\nu_2$ (which includes the
intrinsic thermal and nuclear evolution), and radial reaction
exponents $\zeta_2, \zeta_{\rm L}$ remain constant while the depth
of contact changes. This is only approximately true; and a
self-consistent solution will require numerical integrations. It is
interesting first to look at the implications of equation
(\ref{dotM2power}) when no driving is present, because the solution
is immediate and instructive. This is a situation we encounter in
some large-scale hydrodynamic simulations of mass transfer in
polytropic binaries \citep{DMTF}. Defining a positive dimensionless
mass transfer $y=(-\dot M_2/\dot M_0)^{2/(2n+3)}$, and a
characteristic time scale $\tau=M_2/\dot M_0$, eq.
(\ref{dotM2power}) becomes
\begin{equation}
\frac{\d y}{\d t} = -\frac{\zeta_2-\zeta_{\rm L}}{\tau} y^{n+3/2}.
\end{equation}
The solution can be easily inverted to yield
\begin{equation}
y(t) = y(0)\left[1+y(0)^{n+1/2}(n+1/2)(\zeta_2-\zeta_{\rm
L})t/\tau\right]^{-\frac{2}{2n+1}}\, , \label{nodrivsol}
\end{equation}
where $y(0)$ is the initial mass transfer rate, normalized as above.
This solution illustrates explicitly the role of $\zeta_2-\zeta_{\rm
L}=2(q_{\rm stable}-q)$. In the stable case, $\zeta_2>\zeta_{\rm
L}$, the mass transfer decays asymptotically to zero over a
characteristic time $\tau_{\rm
chr}=\tau/[(n+1/2)y(0)^{n+1/2}(\zeta_2-\zeta_{\rm L})]$, whereas in
the unstable case, $\zeta_2<\zeta_{\rm L}$, it will blow up in
finite time equal to $\tau_{\rm chr}$. Thus the essence of the
stability of mass transfer in a binary is already contained in the
simple case of no driving. The presence of driving exacerbates the
natural instability or, in the stable case, the mass transfer
settles asymptotically to a non-zero stable value. This is what we
observe, for example, in CVs, AM CVns and LMXBs.

Returning now to eq. (\ref{dotM2power}), with the same definitions
as above for $y$ and $\tau$, we obtain for the general case in which
driving is present,
\begin{equation}
\frac{\d y}{\d t} = -\frac{\zeta_2-\zeta_{\rm L}}{\tau}
(y^{n+3/2}-y_{\rm eq}^{n+3/2})\, , \label{dydtpo}
\end{equation}
where $y_{\rm eq}^{n+3/2}\equiv-(\dot M_2)_{\rm eq}/\dot
M_0=(\nu_2-\nu_{\rm L})\tau/(\zeta_2-\zeta_{\rm L})$ is the
equilibrium value normalized to $\dot M_0$. Note that in the stable
case, this value is positive; while it is negative in the unstable
case. Before we attempt to solve the above differential equation, it
is clear from its form and the signs just discussed, that it
describes a stable solution in which $y\rightarrow y_{\rm eq}$ when
$q<q_{\rm stable}$. If, however, $q>q_{\rm stable}$, the r.h.s is
positive even if the mass transfer vanishes initially, and it just
gets bigger as the mass transfer grows. Since $y$ diverges for the
no-driving case in a finite time, the driven case diverges even
sooner.

In order to obtain an analytic solution to eq. (\ref{dotM2power})
which can be compared to the solution of \citet{WebIb}, it is
necessary to cast it in a slightly different form using the
equilibrium rate defined in eq. (\ref{M2doteq}), and modifying
appropriately the definitions of the integration variable and
characteristic time. With the definitions $y_*\equiv [\dot M_2/(\dot
M_2)_{\rm eq}]^{2/(2n+3)}$, and $1/\tau_*\equiv (\nu_2-\nu_{\rm L})
(\dot M_2/\vert(\dot M_2)_{\rm eq}\vert)^{2/(2n+3)}$, the
differential equation for the evolution of mass transfer becomes
\begin{equation}
\frac{\d y_*}{\d t} = \sgn(y_*)\frac{1}{\tau_*} (1- \sgn(y_*)\vert
y_*\vert^{n+3/2})\, ,
\end{equation}
where $\sgn(y_*)$ is the sign of $y_*$. Thus, for the stable case
$y_*>0$, while $y_*<0$ for the unstable case, and $\tau_*$ is
defined positive. The general analytic solution comprising both the
stable and the unstable case can be given in terms of the
hypergeometric function, as follows
\begin{equation}
\frac{t}{\tau_*} = y_*{}_2F_1(1,
\frac{1}{n+3/2};1+\frac{1}{n+3/2},\sgn(y_*)\vert y_*\vert^{n+3/2})\,
. \label{genanalsol}
\end{equation}
While this solution can be easily plotted numerically, it is not
possible in general to invert it to obtain $y_*(t)$. In a few cases,
simpler analytic forms can be obtained. For example, the case
$n=3/2$ yields the solution of \citet{WebIb}, and the case $n=1/2$
is particularly simple:
\begin{eqnarray}
y_* &=& -\tan{(t/\tau_*)} \qquad y_*<0 \qquad {\rm unstable}\cr y_*
&=& \tanh{(t/\tau_*)}\ \qquad y_*>0 \qquad {\rm stable}
\end{eqnarray}

In the case of isothermal atmospheres, the mass transfer rate
\citep{Ri88} is given by :
\begin{equation}
\dot M_2  = - \dot M_0 ~ e^{(R_2 - R_{\rm L})/H}
\end{equation}
where, $H$ is the scale height. This form of the mass transfer
equation is much simpler to integrate than the one for polytropes
considered above. With the same approximations and notation as in
the steps leading to eq. (\ref{dydtpo}), and defining $y=-\dot
M_2/\dot M_0 = \exp{((R_2-R_{\rm L})/H)}$, we obtain
\begin{equation}
\frac{1}{y} \frac{dy}{dt}=  - \frac{\zeta_2 - \zeta_{\rm
L}}{\tau}\frac{R_2}{H} (y-y_{\rm eq})\, . \label{dydtis}
\end{equation}
This can be easily integrated to obtain:
\begin{equation}
y =  \frac{y_{\rm eq}}{1-(1- y_{\rm eq}/y_0) e^{-t/\tau_{\rm iso}}}
\label{yiso}
\end{equation}
where $\tau_{\rm iso} \equiv H/R_2 (\nu_2-\nu_{\rm L})$ is the
timescale required for the driving to change the depth of contact by
$\sim H$, and $y_0$ is the initial value, always positive for
physically meaningful cases. In the stable case $y_{\rm eq}>0$, and
$y\rightarrow y_{\rm eq}$, while $y_{\rm eq}<0$ for the unstable
case and $y$ diverges in a finite time $t_{\rm div} = \tau_{\rm iso}
\ln{(1-y_{\rm eq}/y_0)}$. If no driving is present, we may set
$y_{\rm eq}=0$ and integrate eq. (\ref{dydtis}) for an isothermal
donor. The result is again simple and instructive
\begin{equation}
y = \frac{y_0}{1+(\zeta_2-\zeta_{\rm L}) y_0 \frac{R_2}{H}
\frac{t}{\tau} }\, .
\end{equation}
In the stable case, for any initial mass transfer, the system will
detach and mass transfer will tend to zero. In the unstable case,
any non-zero initial mass transfer will grow and diverge in a finite
time.

\section{Numerical Integration Results} \label{evol}

We now relax some of the constraints imposed in the previous section
and integrate the evolution equations allowing the binary parameters
to adjust self-consistently. Specifically, we compute the changes in
the masses of the components (assuming conservative mass transfer),
allow the binary separation to change as a result of any driving
present, and compute the values of $\zeta_2$ and $\zeta_{\rm L}$ as
they evolve. The values of $\zeta_2$ depend on the adopted
mass-radius relationship for the donor. Here we use Eggleton's
interpolated zero-temperature mass-radius relationship cited by
\citet{VeRa} and \citet{Maet04}, which is a good approximation for
systems containing old, cold white dwarfs. It is possible that the
systems emerge from a common envelope evolution with a massive
hydrogen atmosphere around the donor \citep{DAnt06}, or that the
donor has a finite entropy such that $\zeta_2$ deviates
significantly from the value obtained from the zero temperature
mass-radius relationship for white dwarfs \citep{DelBil05}. In such
situations, the radial variation rate $\nu_2$ is non-zero, and in
fact can reduce the \textit{net driving rate} $\nu_L-\nu_2$, leading
to a lower value for the equilibrium mass transfer (see Eq.
\ref{M2doteq}). Also, a mass radius exponent ($\zeta_2$)
significantly different from $\approx -1/3$ clearly affects the
stability and evolution of the systems at the onset of mass
transfer, and can lead to shrinking orbits even if the mass transfer
is stable with $\ddot M_2\approx 0$. In our subsequent analysis,
where we are concerned about the long term integrations of the OAE,
we set $\nu_2$ = 0 and use the zero-temperature mass-radius
relationship. We note that for the study of the onset of mass
transfer in finite entropy systems like the ones addressed by
\cite{DelBil05} \& \cite{DAnt06}, a more realistic model for the
donor is required.
\begin{figure}[!t]
\centering \epsscale{0.9}\plotone{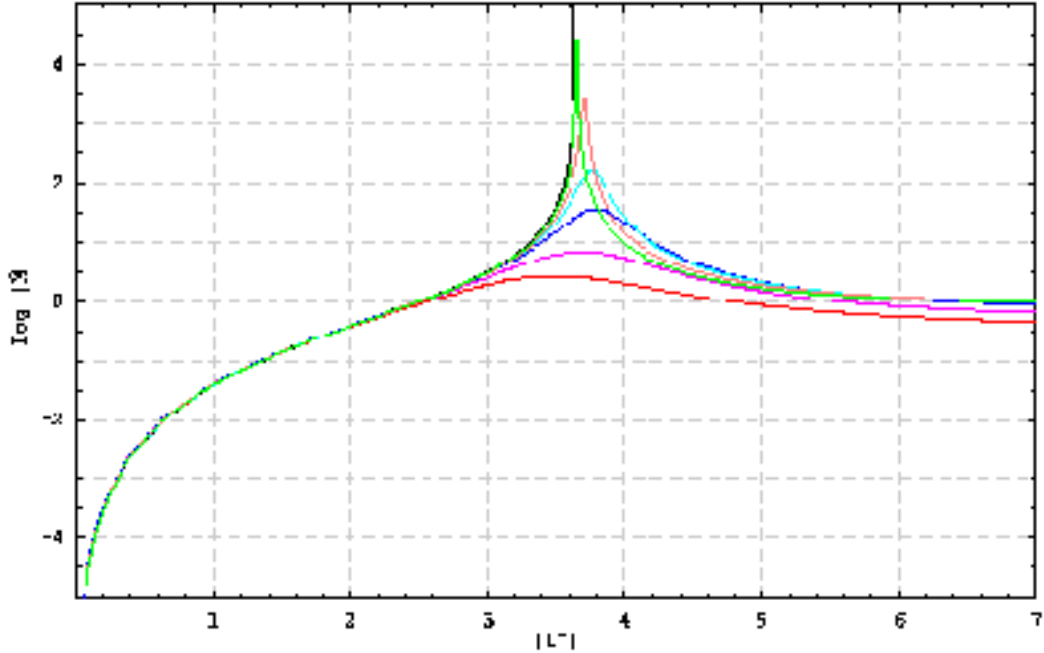} \figcaption{Comparison of
integrations with the analytic solution by Webbink and Iben. The
mass transfer rate normalized to the {\em initial} equilibrium rate
as a function of time in units of the {\em initial} $\tau$: Analytic
(black curve) and numerical -- green ($q=0.663$), orange (0.613),
cyan (0.563 same as WI), blue (0.543), magenta (0.523) and red
(0.513). \label{WIcomp}}
\end{figure}
In order to calculate $\zeta_{\rm L}$, we need to either assume or
determine from other assumptions how the mass and angular momentum
are re-distributed in the binary during mass transfer. As we have
seen in \S\ref{masstran}, this depends on the mode of accretion
appropriate for the binary considered: does the stream impact the
accretor or is an accretion disk present; is the mass transfer
sub-Eddington and conservative, or are mass and angular momentum
being ejected from the system following super-Eddington mass
transfer. For most of the numerical integrations that follow, we use
the appropriate rate of driving by gravitational wave radiation
$\nu_{\rm L}$, wherever necessary we assume a constant driving rate.
However, if one is interested in the relatively rapid phases of mass
transfer that follow contact and onset, then the qualitative
properties of our integrated evolutions do not depend strongly on
these assumptions.

\subsection{Surviving unstable mass transfer}
\label{surviving}

We integrate numerically the evolutionary equations for a sample of
double degenerate binaries whose mass ratios at the onset of mass
transfer exceed the stable value. In order to mimic the conditions
assumed by \citet{WebIb} in their pioneering analysis, we assume a
constant rate of driving, that no mass loss from the system occurs
even during super-Eddington phases, and that a tidally truncated
accretion disk is present at all stages. For these choices, $q_{\rm
stable}\approx 0.49$, and thus we expect unstable mass transfer in
binaries whose initial mass ratio exceeds this value. In Fig. 1 we
present the evolution of mass transfer for a selection of unstable
binaries undergoing conservative disk accretion, including the
specific value $q=0.563$ used by Webbink \& Iben (hereafter WI) to
illustrate their analytic solution. These integrations show that the
mass transfer in an initially unstable binary grows at first
rapidly, peaks, and then evolves asymptotically toward an
equilibrium rate which is also evolving as $q$ changes. The analytic
WI solution diverges in a finite time, while all numerical solutions
reach a peak and then return to stability. The peak transfer
decreases as the initial $q$ approaches $q_{\rm stable}$. The
equations we integrate are orbit-averaged evolution equations (OAE)
in the sense that the rates of change of the orbital parameters are
averaged over one orbital period. This approximation is valid as
long as the evolution is not too rapid, and the eccentricity of the
orbit remains negligible. Furthermore, our formulation does not
include the full effects of tidal distortion and instability
discussed by LRS. Our results suggest that an initially unstable
binary may survive the onset of unstable mass transfer as long as
the mass transfer does not get too big and that the separation does
not get too small. See \S\ref{supercrit} for a discussion of the
limits of validity of the OAE under unstable mass transfer.

\begin{figure}[!t]
\centering \epsscale{0.9}\plotone{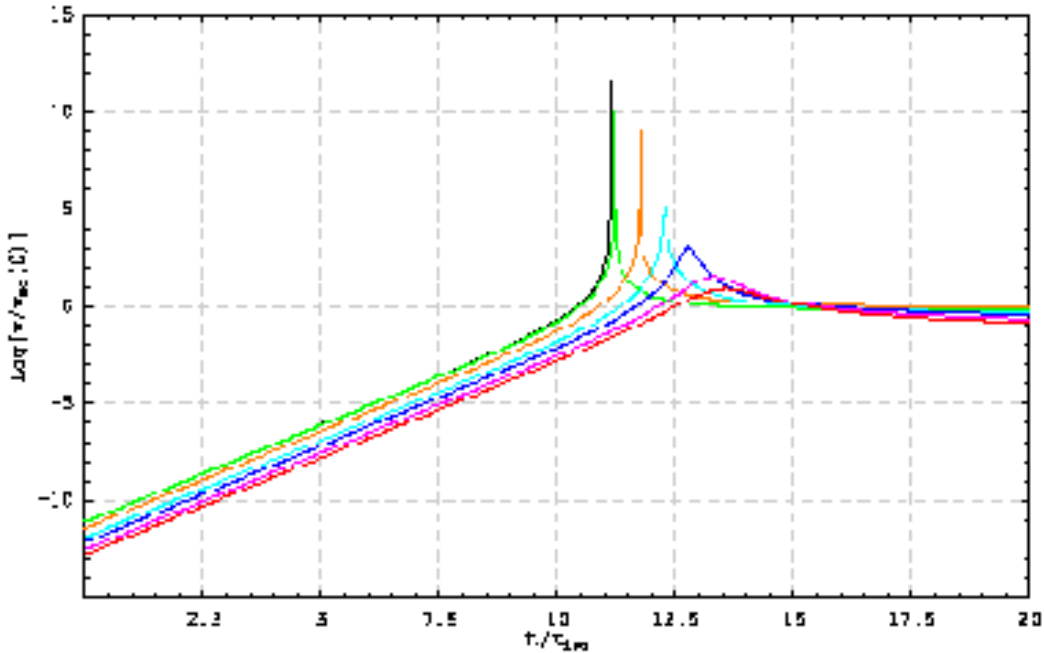} \figcaption{Comparison of
integrations with the unstable isothermal analytic solution. The
natural logarithm of the mass transfer rate normalized to the {\em
initial} equilibrium rate for the case with $q=0.663$, is shown as a
function of time in units of $\tau_{\rm iso}$: Analytic (black
curve) and numerical -- green ($q=0.663$), orange (0.613), cyan
(0.563 same as WI), blue (0.543), magenta (0.523) and red (0.513).
\label{isocomp}}
\end{figure}

It is also interesting to compare the analytic solution given by eq.
(\ref{yiso}) for an isothermal donor in the unstable case, with
numerical integrations of the OAEs for various initial mass ratios.
This comparison is shown in Fig. \ref{isocomp}, with the analytic
unstable solution plotted as the single divergent black solid line.
We elected to plot the natural logarithm of $y$, which is simply the
depth of contact $R_2 - R_{\rm L}$ in units of the pressure scale
height $H$. All the integrations were started from the same initial
depth of contact corresponding to an initial mass transfer of
$10^{-5}$ of the reference rate.
\subsection{Super-Eddington mass transfer}
\label{supercrit}
Another effect may come into play as discussed by \citet{HaWe} when
the mass transfer exceeds the critical Eddington rate and the evolution
becomes non-conservative. We can
incorporate this effect into our numerical integrations, allowing
the excess mass to be blown out of the binary as a wind, carrying away a specific
angular momentum $j_{\rm w}$. We calculate the accreted fraction $\beta$ following
\citet{HaWe}, and modify the evolution equations as follows:
\begin{eqnarray}
\dot J_1 &=& -\beta\dot M_2 j_{1} + \dot J_{\rm 1, tid}\ , \label{J1dotb}
\\
\dot J_2 &=& \dot M_2 j_{2} + \dot J_{\rm 2, tid}\ , \label{J2dotb}
\\
\dot J_{\rm orb} &=& \dot J_{\rm sys} - \left(- \dot M_2[\beta j_1-j_2 +(1-\beta)j_{\rm w}]
+ \dot J_{\rm 1, tid}+ \dot J_{\rm 2, tid}\right)\ , \label{JorbDotb}
\\
\frac{\dot a}{2a} &=& \frac{\dot J_{\rm orb}}{J_{\rm orb}} - \frac{
\dot M_2}{M_2} \left(1 - \beta q - \frac{1-\beta}{2(1+q)}\right)\ . \label{adotb}
\end{eqnarray}
During the direct impact phase, it is reasonable to assume that the
material blown out carries away the characteristic specific angular
momentum of the stream $j_{\rm w}=j_1$. After a disk forms, the
angular momentum advected by the wind will depend on details of the
flow beyond the scope of the present study. For example, if the
material is lost mostly from the vicinity of the accretor, it will
carry approximately the specific orbital angular momentum of the
accretor. Since we do not know exactly how $j_{\rm w}$ varies, we
set $j_{\rm w}=j_1$ throughout. In this case eq. (\ref{JorbDotb})
simply reduces to eq. (\ref{JorbDot}). While it would be possible to
cast the evolution equation for the binary separation (\ref{adotb})
in the same form as eq. (\ref{qa}) by defining
\begin{equation}
q_a \equiv 1+ (1-\beta)q -\frac{1-\beta}{2(1+q)}
- M_2\frac{\beta j_1-j_2+(1-\beta)j_{\rm w}}{J_{\rm orb}}\ , \label{qadefb}
\end{equation}
the explicit appearance of $q$ above makes it obvious that $q_a$
must be calculated self-consistently during the evolution. Note that
when $\beta=1$, the above expression reduces to eq. (\ref{qadef}),
as it should.
\begin{figure}[!t]
\centering \epsscale{0.7}\plottwo{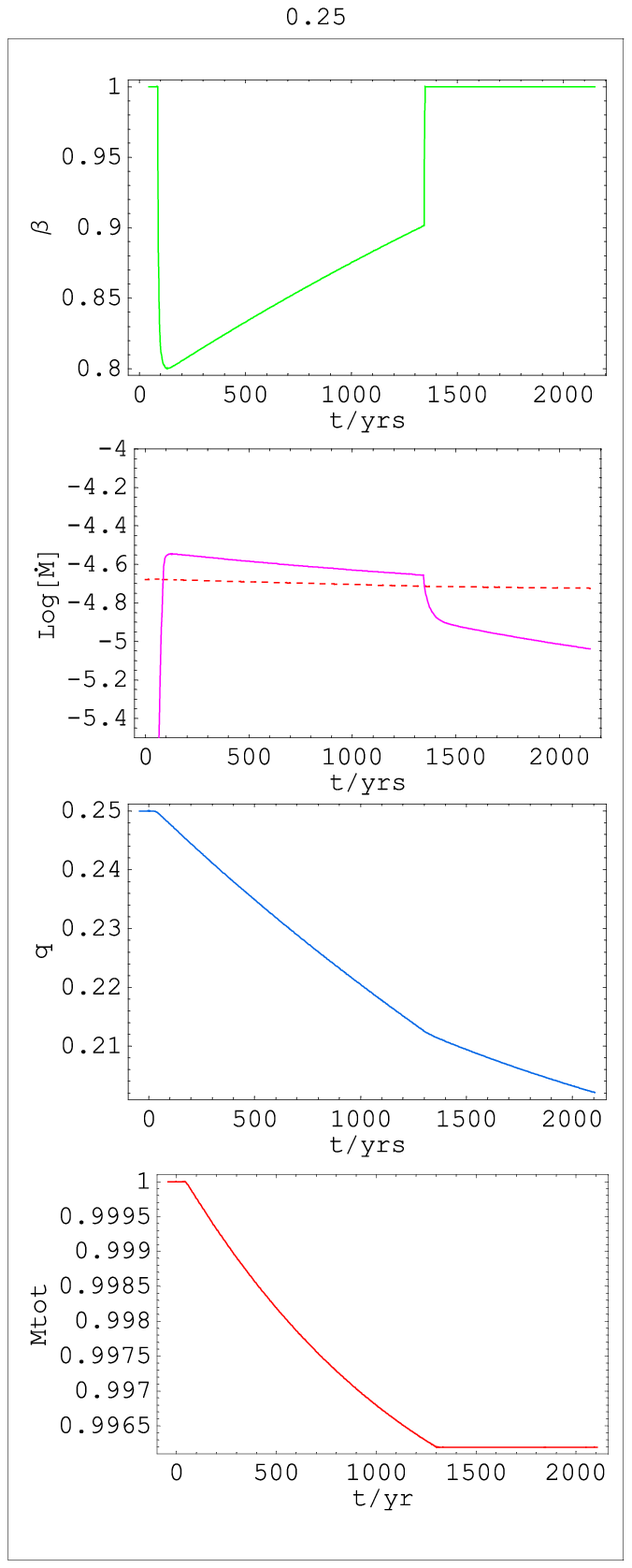}{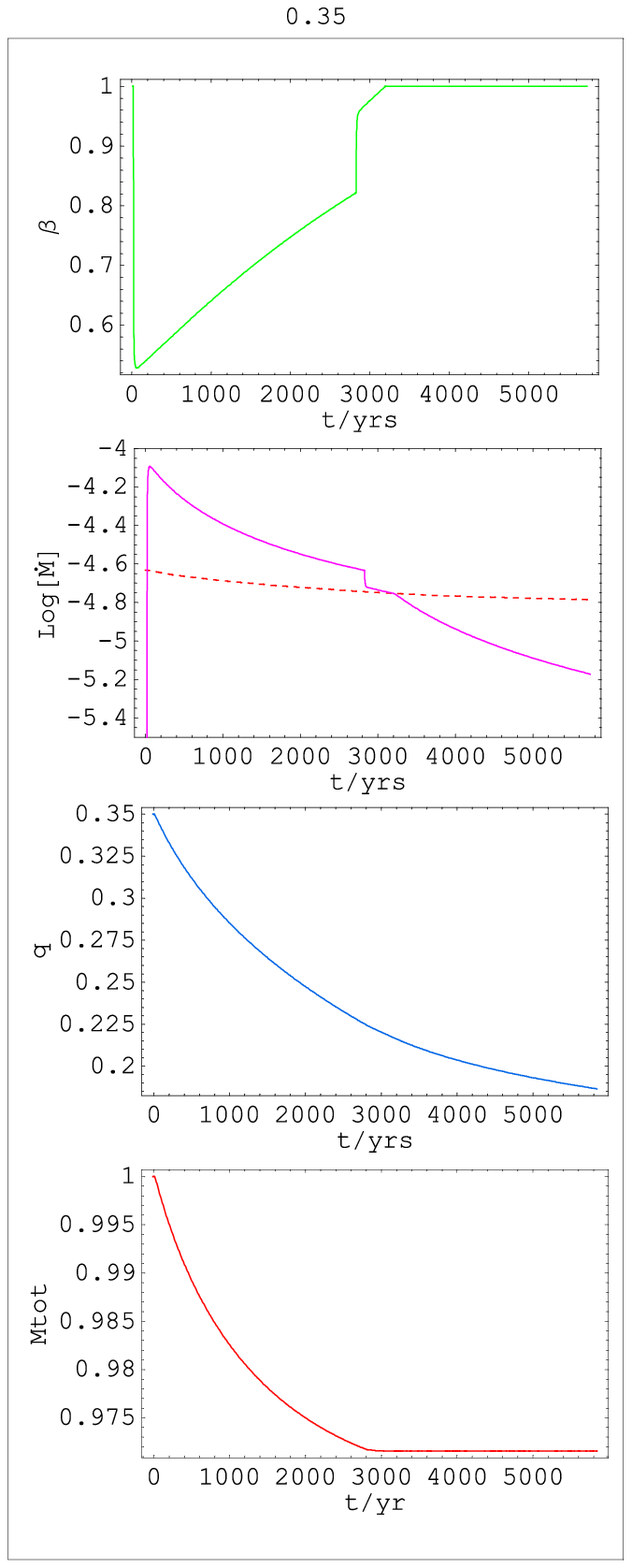} \figcaption{Various
parameters for super-Eddington accretion with direct impact with $q
= 0.25$, just above $q_{\rm stable}$ and $q=0.35$, well above
$q_{\rm stable}$. The panels show from the top down: the accreted
fraction $\beta$; the logarithm of the mass transfer rate (magenta)
and the corresponding Eddington rate (dashed red) in $M_\odot$/yr;
the mass ratio $q$; and the total mass normalized to the initial
value. The abscissa shows times in years from the time of first
contact. The lower $q$ accretes at super-Eddington rates for less
time as compared to the higher $q$; and it does so more gradually
losing less mass (last panel). Even for the initially more unstable
mass transfer (larger $q$), the fraction of mass lost is below 3\%.
\label{evexa}}
\end{figure}
Fig. \ref{evexa} shows two examples of evolutions with a
super-Eddington mass-transfer phase. Because the OAE do not include
tidal distortions of the components or dissipative effects, arising
for example from friction during a common envelope phase, they
always predict survival, no matter how high the mass transfer gets
during an unstable phase. The only exception to this rule occurs if
during the evolution the binary separation falls below the value
$a_{\rm min}$ at which the angular momentum and the energy of the
binary reach a minimum. In that case, further loss of angular
momentum inevitably breaks the synchronism and the system is
secularly unstable. As the binary frequency increases, the spin
frequencies of the components fall behind, tidal synchronization
torques further reduce the orbital angular momentum, and finally a
dynamical instability leads to a rapid merger. However, for DWD
binaries the lower mass component almost always fills its Roche lobe
at a ``contact separation" $a_{\rm c}$ well before this minimum is
reached. Then mass transfer commences and soon the second term in
eq. (\ref{adot}) rises enough to drive the binary apart saving it
from a merger. If initially $q>q_{\rm stable}$ mass transfer rises
even more quickly and causes the binary to expand and recover from
the instability. Clearly the OAE break down if during the transient
the transfer is so high that a significant fraction of the donor
overflows in one orbit, and the dissipative effects mentioned above
may promote a merger even before the mass transfer gets that high.

Suppose that during the transient, the mass transfer rises to $\xi$ times
the critical Eddington rate.
We can estimate the fraction of the donor mass transferred in a single orbit
using the standard assumptions about the donor filling its Roche lobe, and
taking for simplicity a mass-radius relationship $R_i \approx 5\times 10^8
(M_\odot/M_i)^{1/3}$ cm. We find
\begin{equation}
\frac{-\dot M_2 P}{M_2}=\frac{P}{\tau}\approx 10^{-9} \xi
\left(\frac{M_1}{0.5 M_\odot}\right)^{-1/3}\left(\frac{M_2}{0.1 M_\odot}\right)^{-2}\, .
\label{fracc}
\end{equation}
In most of the parameter space this fraction is so small that we can
trust the OAEs to describe approximately the correct behavior within
the limits of the physical effects included in their derivation.
Looking at Fig. 3 of \citet{HaWe}, we see that $\xi\la 10$ for most
cases with the exception of very rare binaries in which the accretor
is already very close to the Chandrasekhar limit. Thus we expect a
certain fraction of the binaries that come into contact with
$q>q_{\rm stable}$ to survive unstable mass transfer, even if it
happens to be super-Eddington. However, to estimate the fraction
that actually make it through, one would have to deal with the
common envelope phase which is beyond the aims of the present study.

\section{Applications}
In the above sections we have developed the basic framework for
investigating the evolution of close DWD systems. By imposing the
appropriate constraints on the OAE, we have compared our OAE
integrations with analytic solutions illustrating the limitations of
the latter. In what follows we investigate the consequences of the
OAE when they are applied with all effects included. One immediate
consequence of the OAE is the phenomenon of tidally induced
oscillations which we describe first. Next, we apply the OAE to a
grid of systems with different initial component masses in the $M_2
- M_1$ space and study the evolutionary outcome of each of these
systems. Finally, we follow the evolution of a single system in
order to achieve a better qualitative understanding of the results
obtained for this system using a full 3-D hydro-code.
\begin{figure}[!t]
\centering \epsscale{0.6} \plotone{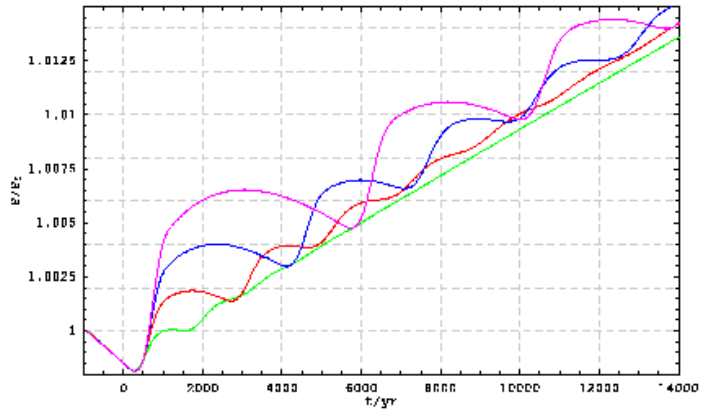} \figcaption{Evolution of the
period normalized to the period of first contact and onset of mass
transfer, for a binary with initial values $q = 0.28$, and
$M_2=0.125 M_\odot$. This choice yields an orbital period of
$\approx300$ s at onset.  The curves shown correspond to different
{\it initial} tidal synchronization times; $\tau_{\rm s_1}$ = 500
yrs (green), 1000 yrs (red), 1500 yrs (blue) and 2000 yrs (magenta).
Tidal timescales are evolved according to eq. (\ref{tidal}). The
longer the tidal timescale, there are more oscillations with higher
amplitude. \label{evcyc1}}
\end{figure}
\subsection{Tidally induced cycles}
\label{cycles} 
\begin{figure}[!t]
\centering \epsscale{1} \plottwo{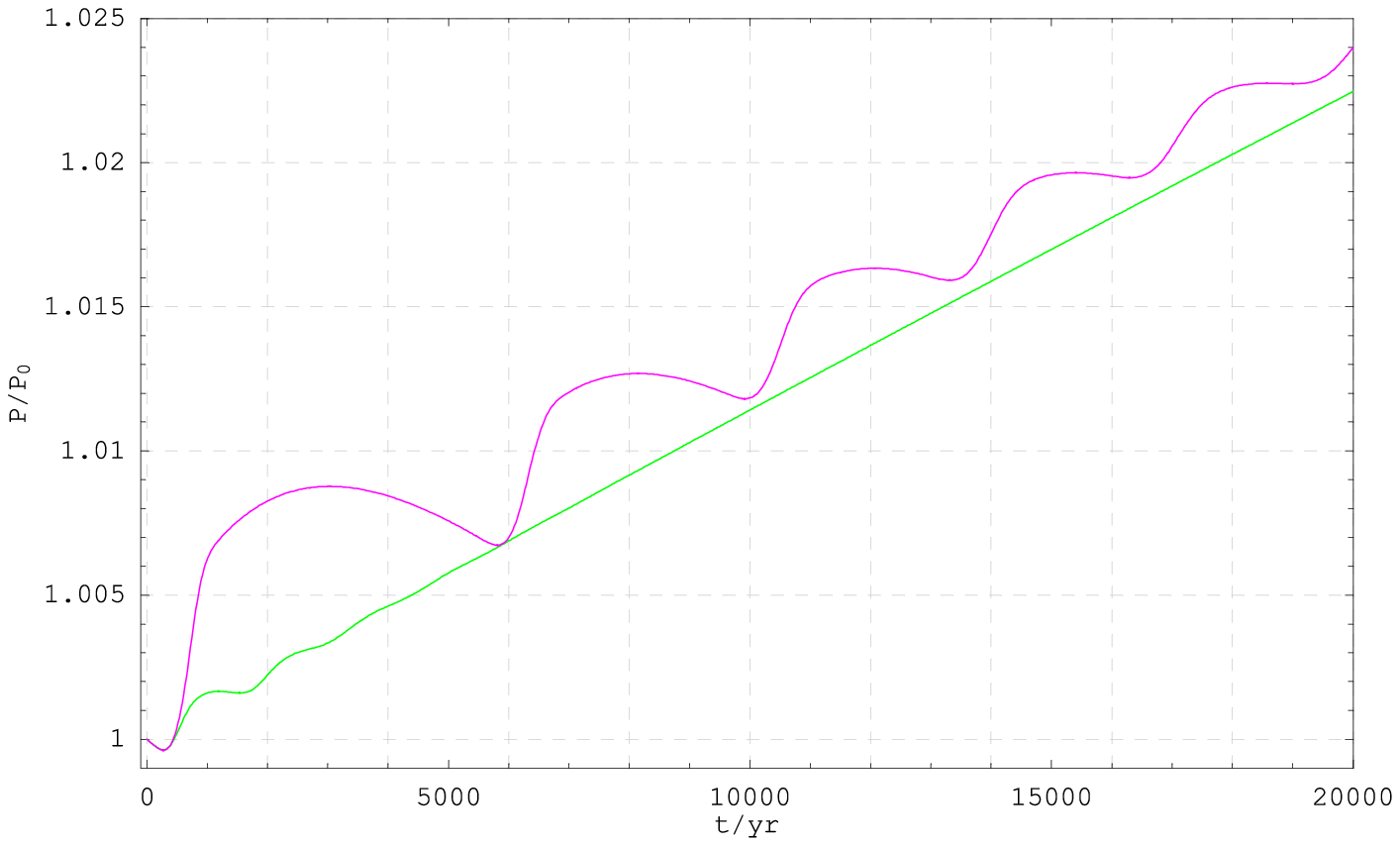}{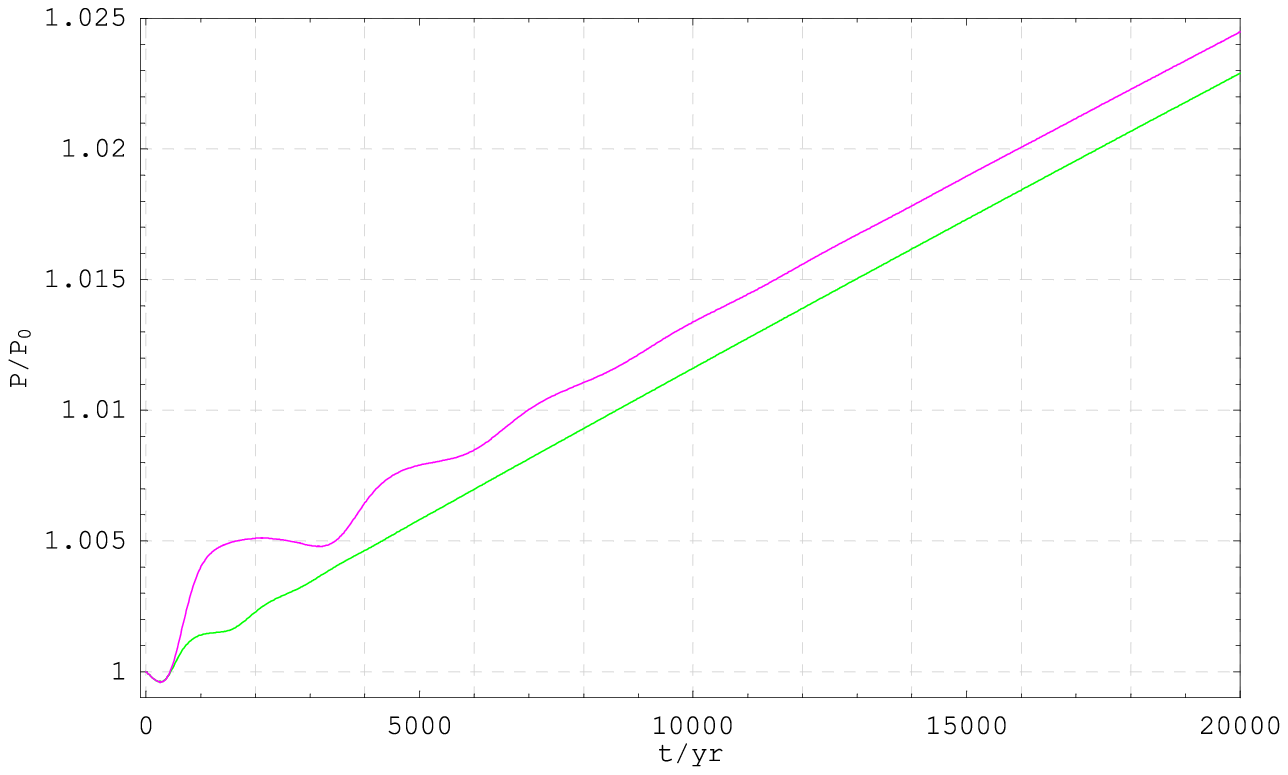}\figcaption{\textit{Left
Panel}: Evolution of a binary with $q = 0.28$ with $\tau_{\rm s_1}$
= 2500 yrs (Magenta) and 500 yrs (Green). \textit{Right Panel}: Same
for $q = 0.24$; $\tau_{\rm s_1}$ = 2500 yrs (Magenta) and 500 yrs
(Green). The time is set to 0 when the binary gets into contact the
first time. The higher the mass ratio, the system goes through a
larger number of oscillations which are of greater
amplitude.\label{evcyc2}}
\end{figure}

As a system gets into contact, the binary separation
decreases at first until the mass transfer rate is high enough to
reverse the trend in $a$. In unstable or near unstable cases, during
this phase the mass transfer timescale $\tau_{\rm M_2}$ decreases
rapidly and becomes shorter than the synchronization timescale of
the accretor $\tau_{\rm s_1}$, allowing efficient spin-up and
building a large asynchronism. As the separation increases, the mass
transfer rate begins to fall and correspondingly $\tau_{\rm M_2}$
increases rapidly (See Fig. \ref{WIcomp}). The synchronization
timescale does not evolve as rapidly as the mass transfer rate and
eventually $\tau_{M_2}\ga\tau_{\rm s_1}$. Now the angular momentum
stored in the spin of the accretor is efficiently returned to the
orbit. If enough asynchronism has been built up in the accretor
during the spin-up phase, the additional injection of spin angular
momentum to the orbit can cause the separation -- and consequently
the Roche lobe radius of the donor -- to increase at a rate faster
than the radius of the donor. On the other hand, the donor radius
increases at a slower rate due to the decreasing mass transfer rate.
This leads to detachment -- the radius of the star cannot keep up
with the increase in the Roche lobe radius. In general, whenever the
effective driving $\nu_L -\nu_2>$ 0, the systems detach. Eventually,
when tides synchronize the spins again, the driving will shrink the
binary back into contact and accretion recommences, slowly at first,
and accelerating as the separation $a$ shrinks and the cycle
repeats. However, as the cycles repeat, the deviations from the
equilibrium rate decrease until the system settles to a steadily
expanding behavior with the mass transfer following the equilibrium
rate closely. In the absence of tidal effects, one would expect the
orbital period to increase monotonically with time once equilibrium
mass transfer has been established for AM CVn type systems. However,
in the presence of appropriate tidal coupling, the systems may
detach and this leads to oscillations in the orbital period --
increasing when the system gets into sufficient contact, and
decreasing when it detaches and the GWR dominates over the tidal
terms. This behavior is evident from Figs. \ref{evcyc1} and
\ref{evcyc2}, in which we plot the orbital period of the binary as a
function of time for different mass ratios and tidal synchronization
timescales. Notice that the number of oscillations the system goes
through is a function of both the tidal timescale and the mass
ratio. For a given tidal timescale, the higher the mass ratio, the
larger the number of oscillations a system is likely to go through.
This is because the higher mass ratio implies that the system is
more unstable which leads to a higher mass transfer rate and thus a
higher degree of asynchronism between the accretor and the orbit. On
the other hand, for a given mass ratio, a longer tidal timescale
allows for a higher spin-up of the accretor, leading to more
oscillations and higher amplitudes. There are, however, limits to
how high or low the tidal timescale (as compared to the mass
transfer timescale) can be in order to observe this behavior. If the
timescale is too high, the spins and orbit are affectively decoupled
whilst if the timescale is too low, the coupling is too efficient to
allow for any significant asynchronism. Thus, in these extremes we
do not observe any oscillations.

During the evolution of the binary, the tidal synchronization times
change because the masses and stellar radii relative to the orbital
separation are changing. We assume that the synchronization
timescales evolve as in \citet{Camp84}:
\begin{eqnarray}
\tau_{\rm s_1} &\propto& \left({M_1\over M_2}\right)^2\left({a\over R_1}\right)^6\cr
\tau_{\rm s_2} &\propto& \left({M_2\over M_1}\right)^2\left({a\over R_2}\right)^6
\label{tidal}
\end{eqnarray}
and choose a normalization ($\tau_{s_0}$) that yields the desired
initial timescales.
As has been mentioned above, tidally induced detachment has
implications for ultra-compact DWD systems \footnote{The higher the
donor mass, the shorter the period at initial contact. Also, a
higher donor mass makes it more likely that the system has an
unstable mass ratio. Thus, in general, oscillations are more likely
in short period systems.}; in particular RX J0806 and V407. In these
systems it is observed that the orbital period is decreasing at a
rate consistent with GWR, but mass transfer is obviously underway in
these systems (\cite{Stroh02} and \cite{Hakala03}). This is at odds
with the theoretical expectation that the orbital period should
increase. Building on ideas discussed in \citet{LaMe}, and
complementing cases considered by \citet{MaNe05}, we have
investigated the possibility of tidally induced detachment. We see
that it is possible for the binary to detach, especially in the case
of unstable, direct impact systems.
\begin{table}[!t]
  \caption{Time spent in different regimes during the oscillation phase as a function of the mass ratio
$q$ and tidal timescale $\tau_{\rm s_1}$.}
  \label{OsctimesTab}
  \begin{center}
    \leavevmode
    \begin{tabular}{lcccccc} \hline \hline
  $q$  & $\tau_{\rm s_1}$ & $T_{osc}$ & $T_1$ & $T_2$ & $T_3$  & $ N_{osc}$ \\
       & (yrs)            & (yrs)     & (yrs) & (yrs) & (yrs)  &  \\
       \hline
  0.28 & 1000 & 3200 & 1260 & 1000  & 936 & 1 \\
  0.28 & 2500 & 23000 & 4450 & 4125 & 14425 & 4 \\
  0.28 & 5000 & 64000 & 4025 & 11655 & 48320 & 5 \\
  0.26 & 1000 & 0 & 985 & $\forall t > T_1$ & 0 & 0 \\
  0.26 & 2500 & 5400 & 1750 & 810 & 1840 & 1 \\
  0.26 & 5000 & 15200 & 1600 & 4100 & 9510 & 2 \\
  0.24 & 1000 & 0 & 710 & $\forall t > T_1$ & 0 & 0 \\
  0.24 & 2500 & 3600 & 1130 & 1820 & 650 & 1 \\
  0.24 & 5000 & 5500 & 1110 & 2300 & 2100 & 1 \\ \hline
      \end{tabular}
  \end{center}
\end{table}
The system parameters like the mass of the donor, accretor and the
various angular momentum loss mechanisms are not accurately known
for the known AM CVn systems. We have calculated the amount of time
the tidal oscillations are in effect in the case of a DWD system
after it first gets into contact by loss of GWR for different system
parameters. In Table \ref{OsctimesTab} we have tabulated the
relevant timescales as a function of the mass ratio $q$ and the
tidal timescale $\tau_{\rm s_1}$ of the accretor. $T_{\rm osc} = T_1
+ T_2 + T_3$ is the timescale for which the oscillations last after
initial contact. $T_1$ represents the time when the binary is in
contact but the separation is decreasing, $T_2$ is the time when the
system is in contact and the separation is increasing, whilst $T_3$
represents the time for which the system is out-of-contact. $ N_{\rm
osc}$ represents the number of oscillations a system encounters
during its evolution. We see that for a given mass ratio, a system
tends to spend an increasing amount of time out of contact with
increasing tidal timescales. Moreover, the more unstable the mass
ratio, the larger the number of oscillations and the timescale for
which the oscillations last.

A binary can spend a considerable amount of time in which tides are
effective, especially in the case of unstable mass transfer. In
fact, since the systems also spend quite a significant fraction of
time out-of-contact, there should be many more systems with short
periods than can be observed. However, even in the most favorable
case, a given system spends less than 30\% of its time in a regime
where the system is in contact and the orbit is shrinking. Thus it
is unlikely that tidally induced oscillations are responsible for
the observations of $\dot P < 0$ for RX J0806 and V407.
Nevertheless, the probability that we catch a system in contact with
$\dot P < 0$ is enhanced significantly as compared to the case when
there are no oscillations \citep{Kalo05}. For example, for the $q$ =
0.26 case the system does not undergo any oscillations and $T_1 \sim
$ 1000 yrs for $\tau_{\rm s_1}$ = 1000 yrs. A more promising idea is
the recent proposal by \citet{DAnt06}, according to which the
behavior of these systems can be understood if the donor possesses a
substantial non-degenerate atmosphere which allows the donor to
shrink as mass transfer proceeds. Under these conditions mass
transfer tends to be more stable, at least initially, and cycles are
unlikely.

In the next section we follow a large number of evolutions
with different initial conditions by integrating the OAEs for $10^9$ yr,
in order to address the question of which kind of systems, and under
which conditions, are likely to experience the cycles described here.

\subsection{Exploring Evolutionary Outcomes}
\label{popsyn}

We explore the parameter space for DWD binaries by investigating the
different types of binary evolution that can occur under a variety
of initial assumptions. Our procedure consists of selecting the
initial masses for binary components and evolving them under driving
by gravitational radiation for $10^9$ yr. In the most general case
we include tidal coupling between the orbit and both components, we
allow mass loss in super-Eddington cases, and we include the
transition from direct impact to disk accretion. We also choose the
tidal synchronization timescales for either the accretor alone, as
in \citet{Maet04} or for both components allowing even the donor to
become non-synchronous. As a check, we have evolved a grid of models
suppressing the tidal and advective terms from the donor and
choosing the same synchronization timescale as \citet{Maet04}. Under
these assumptions our results are indistinguishable from theirs.

\begin{figure}[!h]
\centering \epsscale{0.7} \plotone{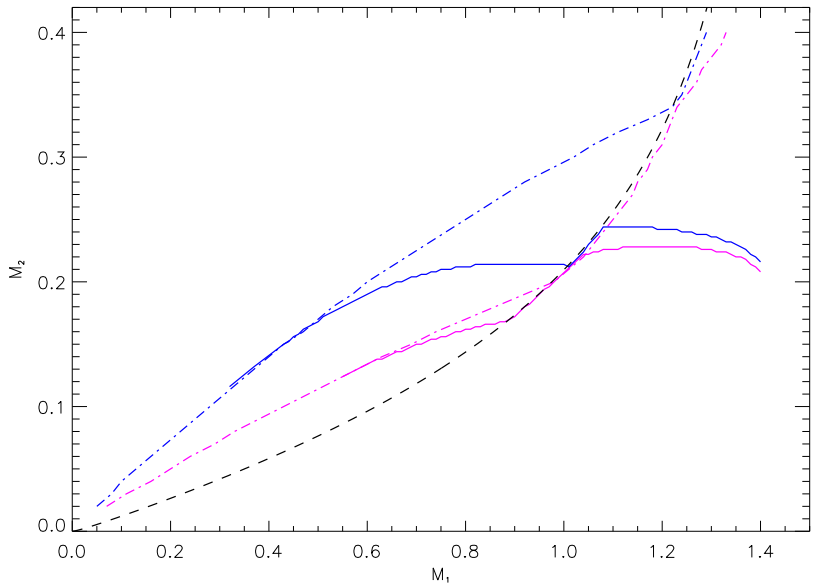}\figcaption{Mass-transfer
stability limits (dash-dot lines) and super (above) and
sub-Eddington (below) accretion boundaries (solid lines) with (blue)
and without (magenta) the donor terms included. The thin dashed
black line divides direct impact systems from disk accretion systems
\citep{MaSt02}. Because the transition from direct impact to disk
accretion makes mass transfer more stable, both the stability limits
and Eddington accretion rate boundaries follow closely the locus of
that transition toward higher donor masses (see text for details).
\label{SuperEexp}}
\end{figure}

Before presenting the results of the evolutions, we investigate the
expected behavior of the systems, focussing on the stability limits
and whether the mass transfer is super-Eddington or not. In Fig.
\ref{SuperEexp} we plot the stability limits for two cases: one with
the donor spin properly accounted for (blue line) and the other with
the donor effects ignored (magenta line). In addition for each case,
we plot the locus of points for which $\dot M_{2_{\rm{eq}}} \approx
\dot M_{\rm{Edd}}$, i.e., the locus of points that defines a
transition from super-Eddington accretion to sub-Eddington
accretion. Note that these curves represent the equilibrium mass
transfer and Eddington rates of the system \textit{at initial
contact} where we assume that the systems are synchronous and thus
the tidal terms are zero.
For direct impact systems the accretion rate is always
super-Eddington if $M_2 >0.21 M_\odot$ (0.17 $M_\odot$ if donor
terms are neglected). This is because of two reasons -- i) in
general, the higher the donor mass, the higher the mass ratio and
thus the systems are closer to instability, which in turn implies a
higher $\dot M_{2_{\rm{eq}}}$ and ii) the higher the accretor mass,
the lower the threshold for super-Eddington accretion \citep{HaWe}.
However, as can be seen from Fig. \ref{SuperEexp}, the transition of
the systems from direct impact to disk leads to changes in the
stability properties of these systems -- they tend to be slightly
more stable because the loss of orbital angular momentum to the
accretor spin in disk systems is smaller than in the direct impact
case. This is reflected in the slight upturn in the stability curve
around $M_1 \sim$ 1 $M_\odot$ (0.85 $M_\odot$ if donor terms are
neglected). The result of this is that the equilibrium mass transfer
rate for the disk systems is lower than it would have been for that
same system if it were a direct impact system. Consequently these
systems tend to undergo sub-Eddington accretion, and the locus of
the transition between super and sub-Eddington accretion systems
follows the stability curve for both cases. However the disk
transition `saves' only a few systems because DWD binaries with $M_2
> 0.25 M_\odot$, (0.23 $M_\odot$ if donor terms are neglected) are
all super-Eddington initially.

\begin{figure}[!h]
\centering \epsscale{1} \plottwo{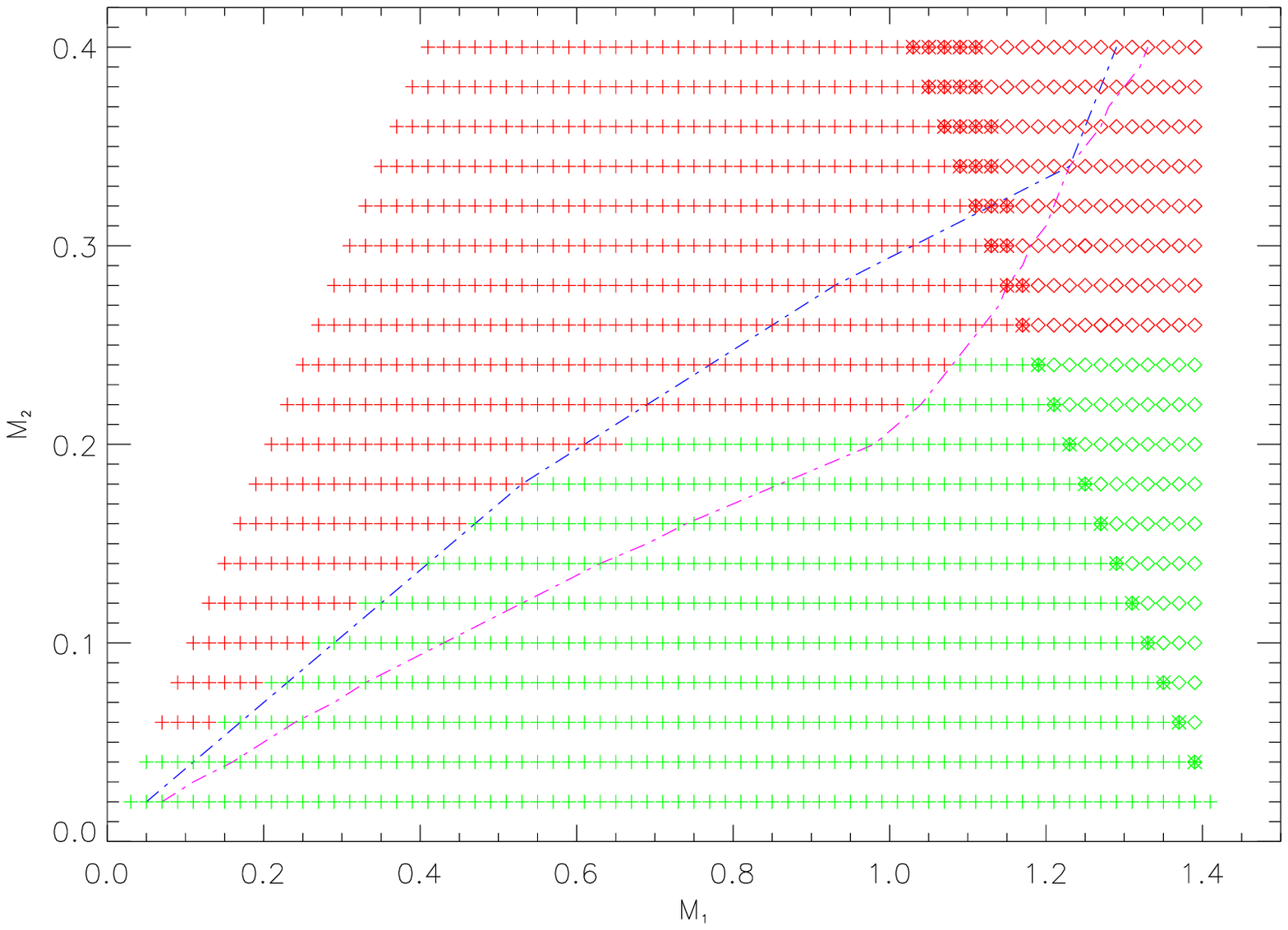}{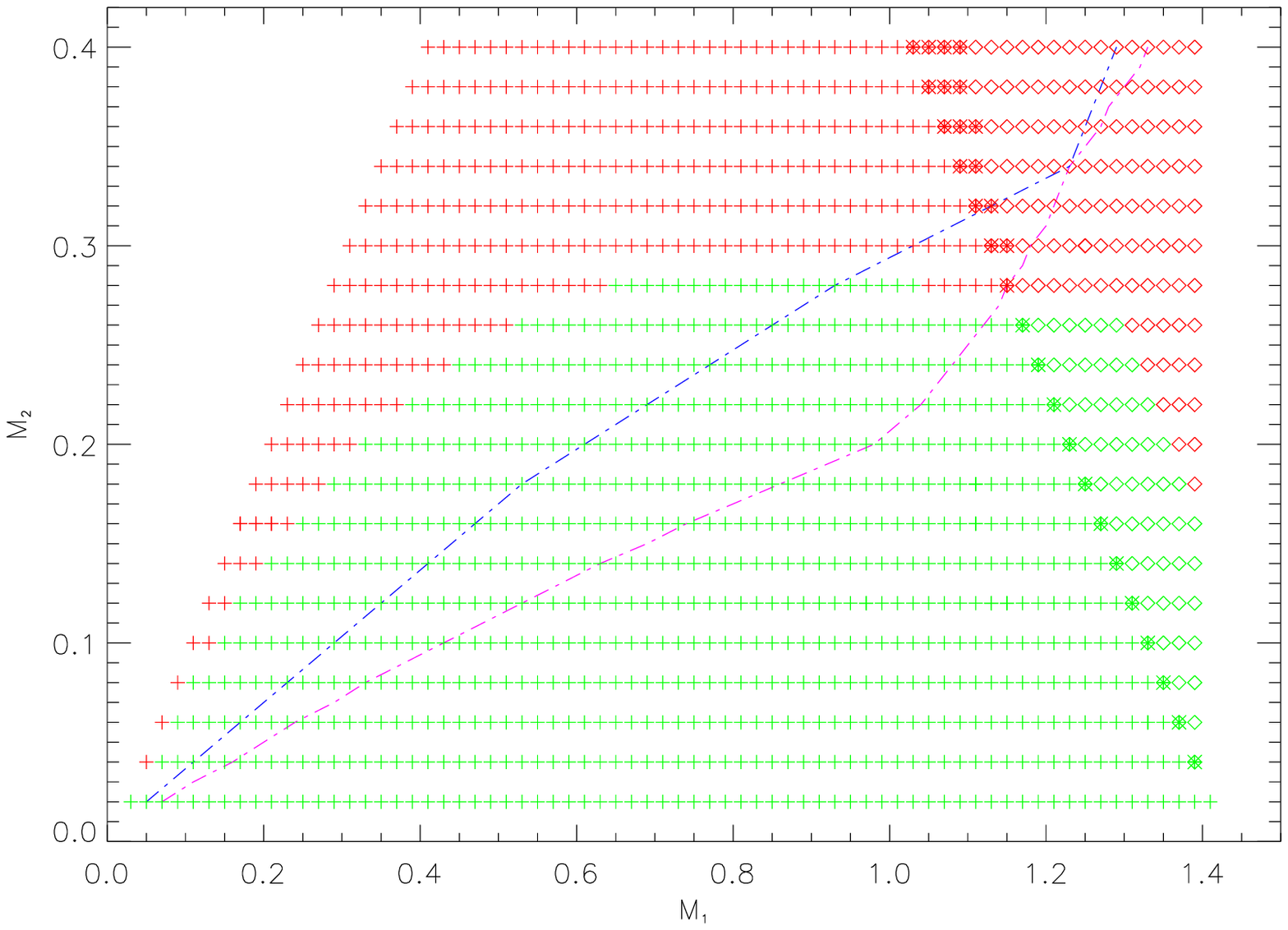} \figcaption{Evolution for
$10^9$ yrs for an initial $\tau_{s_1}$ of $10^{15}$ yrs (left panel)
and for $\tau_{s_1}$ of 10 yrs (right panel). This synchronization
time is for the initial configuration and evolves according to eq.
(\ref{tidal}). $\tau_{s_2}$ is calculated based on whatever
$\tau_{s_0}$ is required to obtain the desired value of
$\tau_{s_1}$. The red symbols represent super-Eddington accretion,
the green are sub-Eddington, the pluses ($+$) and hollow diamonds
($\lozenge$) represent systems with total mass below and above the
Chandrasekhar limit, respectively. Among the latter, asterisks over
diamonds indicate systems in which the accretor does not reach the
Chandrasekhar limit in $10^9$ yr. The blue dash-dot line is the
initial stability boundary with all donor effects included. The
magenta line is the stability boundary without these effects, shown
here for comparison. Note the transition in the super and
sub-Eddington accretion rate around an $M_2$ of 0.2 $M_\odot$ to the
latter stability curve. \label{popsynus}}
\end{figure}

\begin{figure}[!h]
\centering \epsscale{0.7} \plotone{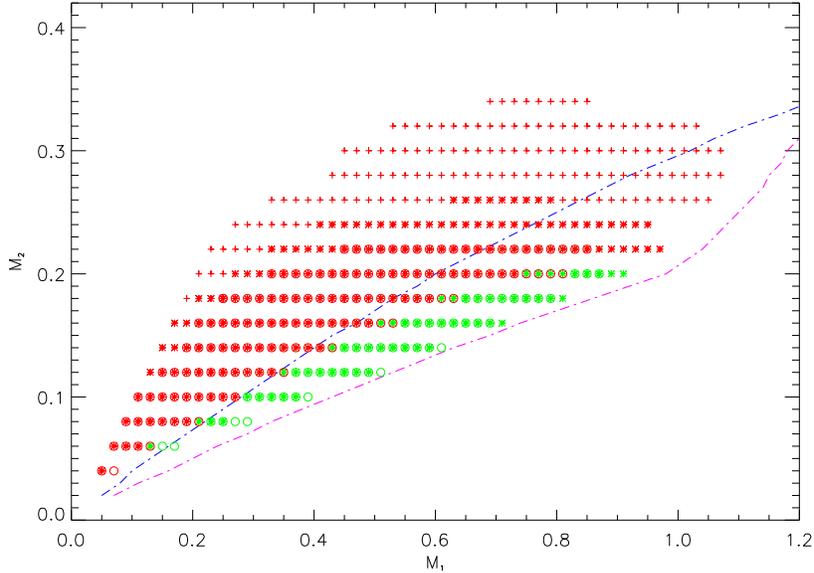} \figcaption{Systems which
undergo `oscillations' at least once during their evolution for
$\tau_{s_1}$ = 500 yrs (pluses, $+$), 2500 yrs (crosses, $\times$)
and 5000 yrs (open circles, $\circ$). Red indicates super-Eddington
transfer during any part of the evolution, whereas green indicates
sub-Eddington transfer throughout the $10^9$ yr evolution. The
magenta and blue lines are the stability limits as in previous
figures. $\tau_{s_2}$ is held at a constant 100 yrs.
\label{popsyncyc}}
\end{figure}

We present now the results of the numerical integrations of the OAE
for a grid of models in the $M_2-M_1$ parameter space. In Fig.
\ref{popsynus}, we plot the results for two extreme values of the
accretor tidal synchronization timescales -- one with an extremely
large tidal timescale ($10^{15}$ yrs) corresponding to inefficient
coupling and one for a short timescale of 10 yrs corresponding to
highly efficient spin-orbit coupling. Firstly, we note a
significantly increased region (as compared to when the donor terms
are ignored) of parameter space corresponding to sub-Eddington mass
transfer and probable survival as a long-term, stable mass transfer
binary of the AM CVn type. Also, from our discussion above, we
expect that the case with inefficient tidal coupling (left panel in
Fig. \ref{popsynus}) to match the curves in Fig. \ref{SuperEexp}.
The transition from super to sub-Eddington accretion overlaps the
stability boundary until $M_2 \approx$ 0.21 $M_\odot$ after which it
follows the stability curve defined by the direct impact to disk
transition, in accordance with our expectations. On the other hand,
since the tides almost always act to stabilize the system by
effectively reducing the driving rate, a simplistic analysis based
on comparing the \textit{initial} equilibrium mass transfer and
Eddington rates has only partial validity. While the threshold for
super-Eddington accretion remains unaffected by the tidal coupling,
the mass transfer rate is significantly lowered due to the reduced
driving rate, and can be below the Eddington accretion rate. Systems
which have super-Eddington accretion rates when the tidal coupling
is inefficient, can thus accrete at sub-Eddington rates if the tidal
coupling is strong. As a consequence we expect the domain of
sub-Eddington accretion to extend to higher donor masses, and on
comparing the left and right panels of Fig. \ref{popsynus} this is
what we observe.
Finally, the locus of systems that undergo oscillations in their
separation and binary period is illustrated in Fig. \ref{popsyncyc}.
Whether a system undergoes oscillations or not, depends primarily on
two factors: the accretor tidal synchronization timescale $\tau_{\rm
s_1}$ and the mass ratio $q$. As can be seen from Fig.
\ref{popsyncyc}, oscillations are seen to occur only in a certain
domain around the stability curve, and that domain decreases with
increasing tidal synchronization times. This is due to the fact
that, for higher donor masses, the mass transfer timescale
$\tau_{M_2}$ is quite short and if the tidal timescale $\tau_{s_1}$
is much longer than $\tau_{M_2}$, the spin and the orbit are
effectively decoupled. As a result there is minimal return of the
spin angular momentum to the orbit and consequently, there are no
oscillations for these long timescales. In fact, referring to Fig.
\ref{popsyncyc}, we observe that systems with high donor mass ($M_2
> 0.35 M_\odot$) do not undergo any oscillations at all. Again, this
is because the mass transfer rates are high in this domain and thus
the tidal timescales (considered here) are much longer than the mass
transfer timescales. Consequently, the radius of the donor keeps up
with the increase in the Roche lobe radius throughout the evolution,
and the systems stay in contact.

On the other hand, systems to the bottom right (the ones with low
donor mass and high accretor mass) are stable systems and are more
likely to be disc-systems. Thus the accretor is not spun up as much
as in the case of less stable or unstable systems. Moreover in the
disc systems, tides are extremely efficient in returning angular
momentum back to the orbit. Both these factors conspire to decrease
the level of asynchronism achieved by the accretor and consequently,
these systems do not undergo oscillations.

We have also studied the effect of the donor's tidal synchronization
timescale on the domain over which systems undergo tidally induced
oscillation cycles. In most cases, the level of the donor's
asynchronism is rather low. Thus the magnitude of the term
associated with the tidal effects of the donor in the driving rate
$\nu_L$ (eq. (\ref{RLdotarr})) is much smaller than the
corresponding term for the accretor. As a result, the donor
synchronization timescale has a limited impact on the domain over
which the systems oscillate.

\subsection{Comparison with hydrodynamic simulations}
\label{hydrocomp} 
\begin{figure}[!h]
\epsscale{.85}\plotone{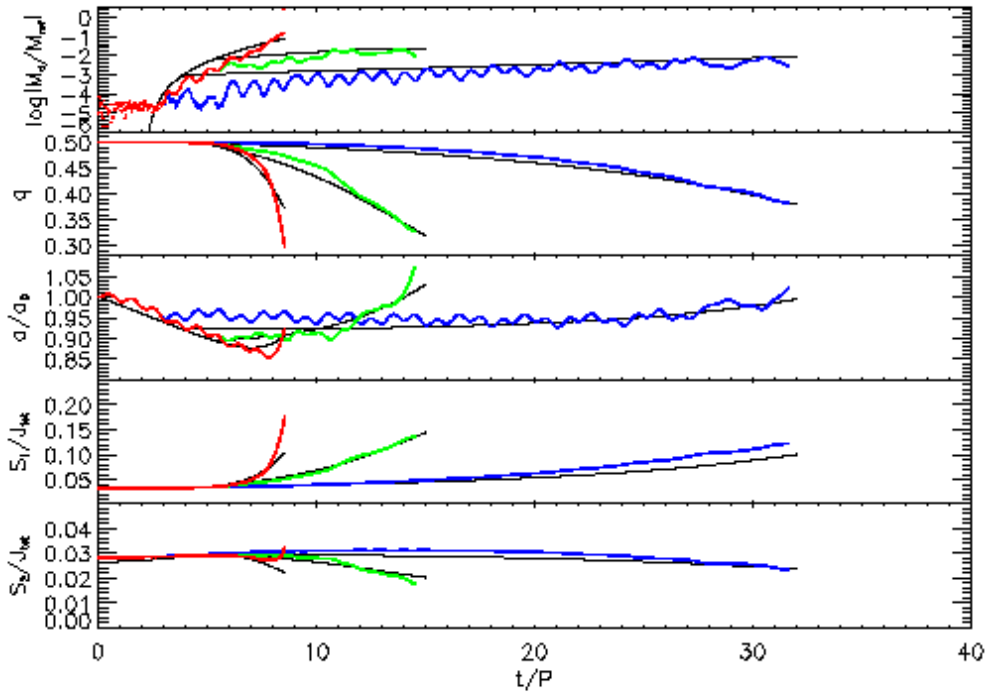} \figcaption{Comparison with some of the
numerical simulations ($q = 0.5$ run) in \citet{DMTF}. Three
simulations were performed for the same initial conditions but the
binary was driven by angular momentum losses at the level of 1\% per
orbit for different times in order to achieve increasing depths of
contact: Q0.5-a (blue; driven initially for 2.7 orbits), Q0.5-b
(green; driven for the first 5.3 orbits) and Q0.5-c (red; driven
throughout). The solid black curves show the accretion rate in donor
masses per period, the binary mass ratio, the separation normalized
to the initial separation, and the spin angular momenta as predicted
by the OAEs, while the dotted color curves show the same quantities
as derived from the results of the simulations. We arbitrarily
change $\tau_{\rm s_i}$ of the donor and accretor to match the
Q0.5-a run and then predict the outcomes of the other simulations.
Here, $\tau_{\rm s_1} \sim 150 P$ and $\tau_{\rm s_2} \sim 3.5 P$
initially.\label{numsim}}
\end{figure}

One of the stated objectives of this paper is to
develop a theoretical framework to interpret and analyze results of
large-scale, self-consistent, 3-D hydrodynamic simulations of
binaries undergoing mass transfer \citep{DMTF}. To this end, we
apply the same initial conditions to our OAE as have been to the
various runs carried out by D'Souza et al. We have the tidal
normalization factor ($\tau_{s_0}$, see eq. (\ref{tidal})) and the
mass transfer rate scaling ($\mathfrak{m}$) -- such that $\dot M_2 =
-\mathfrak{m} \dot M_0 f(\Delta)$ -- as `free parameters', which we
adjust so as to obtain as close a match to the hydro-runs as we can.
In the particular run shown in Fig. \ref{numsim}, $\tau_{s_0}$ =
0.75 and $\mathfrak{m}$ = 35.0. This choice is not unique -- indeed,
we get reasonable `fits' even for other combinations of these `free
parameters'.
It should be noted that in the numerical simulations of \citet{DMTF}
the minimum resolvable mass transfer was on the order of $\sim
10^{-5} M/P$, which translates for short period AM CVn binary
parameters to $\sim 10^4 \dot M_{\rm Edd}$! Also, in the case of the
hydro-runs one observes severe distortion of the accretor and the
formation of an accretion belt around the accretor towards the end
of the evolution. These features cannot be easily incorporated in
the OAEs, and so the later stages of evolution especially in the
case of Q0.5-c, cannot be properly represented in the OAE. An added
complication is that for the hydro-runs, the driving was cut off
after the systems were thought to have reached a deep enough
contact. Since the effective density levels, especially near the
edge of the stars in the 3-D numerical model differs from an ideal
$\rm{n=3/2}$ polytrope due to the finite numerical resolution of the
code, the depth of contact achieved after a certain amount of
driving for a certain period of time is not necessarily the same for
the hydro-runs and the OAE runs. This is especially true for Q0.5-a,
because it is the one which is most sensitive to the depth of
contact at the instant the driving is cut-off. For runs Q0.5-b and
Q0.5-c, the depth achieved is deep enough to make small differences
between the hydro-runs and the OAE unimportant. We are working on
another set of runs in which the driving is not cut off and the
systems are driven at slightly lower (and more realistic) rates. We
hope that this will eliminate another source of discrepancy between
the hydro-runs and the OAE.

Despite the above mentioned shortcomings, the OAEs do reproduce
reasonably well the behavior of the binaries we have studied. One
notices that the hydro-runs have a gentler slope initially which
progressively gets steeper as compared to the OAE runs. This, we
believe, is a consequence of the complicated fluid flow around the
$L_1$ point and the distortion of the donor star. Moreover, during
the initial stages of the evolutions, the 3-D hydrodynamic
simulations are subject to numerical noise which is not the case for
our numerical integrations. The relative significance of this noise
diminishes as the mass transfer rate increases during the evolution.
Also, the epicyclic motion that one encounters in the hydro-runs
(see \cite{DMTF} for details) cannot be reproduced in our results,
since we assume circular orbits. Thus the behavior of the OAE is
much smoother than the numerical hydro-runs with no abrupt changes
in slopes of the various parameters.

We conclude from the above discussion and Fig. \ref{numsim}, that
the OAEs confirm that: a) tidal effects play an important role in
the numerical simulations of the binary, b) direct impact accretion
is an important effect and can lead to significant spin-up of the
accretor at the expense of orbital angular momentum, and c) the OAE
prediction that systems that are initially unstable can indeed
survive mass transfer seems to bear out in the hydro-runs despite
the rather extreme conditions to which the binaries in the
hydro-runs are subjected.

In addition to the Q0.5 runs presented above, \cite{DMTF} also
present other runs which result in the disruption of the binary. As
can be seen from their Fig. 3, 6 and 10, the donor star becomes
increasingly distorted towards the end of these simulations. The
OAE, at least in their present form, are not capable of accounting
for the distortion of the components and the consequence of these
distortions on the fate of the binary. Additional hydro-runs with
different values of the mass ratio $q$ and lower driving rates $\nu$
(though these rates are still orders of magnitude higher than
realistic values) are being carried out \citep{Motletal06}. In
principle this can help, for example, in determining a limit on the
mass ratio $q$ ($>~ q_{\rm stable}$) above which the tidal
disruption of the binary is likely. However, other physical effects
not yet included in the 3-D simulations, such as radiation and the
formation of a common envelope, are more likely to determine the
fate of such systems.

The tidal time scales that most closely account for the behavior
observed in the simulations will serve as a measure of the numerical
dissipation present in the simulations. We will present elsewhere a
more comprehensive study of the numerical dissipation and its
dependence on spatial resolution.

\section{Discussion}
\label{discuss} We have re-examined the standard circular
orbit-averaged equations (OAEs) that describe the evolution of mass
transferring binaries allowing for advective and tidal exchange of
angular momentum between the components. We found that the mass
transfer stream issuing through the $L_1$ point has two effects in
the internal redistribution of angular momentum in the binary: 1) it
delivers a specific angular momentum $j_{\rm circ}$ to the accretor
spinning it up at the expense of the orbital angular momentum; and
2) it reduces the spin angular angular momentum of the donor by a
specific amount $\sim R_2^2\omega_2$, and ultimately couples to the
orbit via tides. In the cases examined in this paper, the effect of
this additional term is mildly stabilizing. For example, with
parameters thought to be appropriate for the two short-period,
direct impact binaries V407 Vul and RX J0806, the additional donor
term is $\sim 0.1$ (see eq. (\ref{qstab})), so that the net effect
of the consequential terms (accretor plus donor) $\approx -0.3$, is
still de-stabilizing but less than estimated by \citet{Maet04}, and
$q_a\approx 0.7$. Its effect on the evolution of other systems
remains to be explored further, but in systems driven by magnetic
braking, it is likely to be relatively minor since it is relatively
smaller and would be masked by the magnetic torques acting directly
on the donor.

We have extended analytically and numerically our understanding of
the evolution of stable and unstable mass transfer in semi-detached
binaries, with special emphasis on DWD binaries. In particular we
have extended the analytic solutions of the type discussed by
\citet{WebIb} to other polytropic indexes and the isothermal case.
The analytic solutions predict that if $q
> q_{\rm stable}$ at contact, the mass transfer rate diverges in a finite
time implying the catastrophic merger of the two components. The
OAEs on the other hand, always predict that a binary undergoing
unstable mass transfer would survive after a phase of rapid mass
transfer which in many cases would reach super-Eddington levels.
This is because, unlike the analytic case, in the numerical
integration of the OAE we allow the binary parameters to evolve
self-consistently throughout the evolution. Thus, even if initially
$q > q_{\rm stable}$, at some point in the evolution $q < q_{\rm
stable}$, and the systems settle to the equilibrium mass transfer
rate corresponding to the current values of the system parameters.
While we do incorporate mass loss due to super-Eddington accretion,
following the treatment along the lines of \citet{HaWe}, clearly the
OAEs must break down if a common envelope forms. The treatment of
common envelope evolution is beyond the scope of the present paper,
but we suspect that many systems previously considered doomed to a
merger would actually survive provided that the mass transfer peak
is not too high. This point clearly needs further investigation and
probably requires 3-D hydrodynamic simulations with the inclusion of
radiative effects.

One interesting consequence of the tidal coupling of the accretor to
the orbit is the appearance of mass transfer oscillation cycles
occurring for a limited time after the onset of mass transfer. The
likelihood of these cycles is higher in situations where the mass
transfer is high, or rises to a high value rapidly. Therefore they
tend to occur near and around the stability boundaries. Note that
all the systems that undergo oscillations are direct impact systems,
at least for the system parameters and synchronization timescales we
have considered. Thus it is highly unlikely that systems that have
accretion disks will undergo the tidally induced oscillation cycles
in the case of DWD binaries.

The presence of a massive ($\sim 0.01 M_\odot$) non-degenerate
atmosphere on the donor at the onset of mass transfer \citep{DAnt06}
can affect the stability and evolution of these systems. For
example, a large and positive $\zeta_2$ would imply $q_{\rm
stable}>1$, and has significant implications for cycles, stability
boundaries and super-Eddington mass transfer. The full consequences
of these circumstances remain to be explored further.

DWD systems are some of the most common compact systems in the
galaxy and are of particular importance for the space based
gravitational wave detector LISA. AM CVn systems are guaranteed
sources for LISA and the knowledge of possible evolutionary
trajectories is valuable. The framework we have outlined in this
paper can be used to generate templates for short period DWD's in
general and AM CVn systems in particular. Similar work has been done
already (see for example, \cite{RavToh06} and \cite{SVN05}); but the
effects of the tidal couplings and the advective term associated
with the donor spin remain to be incorporated into future studies.

The LSU theory group has performed a number of large-scale 3-D
numerical simulations of interacting binaries with polytropic
components \citep{MTF, DMTF}. These simulations did not include
enough physics to tackle the common envelope evolution, but they
should be viewed as the first steps toward that goal. In the
meantime, we have used the OAEs with suitably adjusted tidal
coupling time scales to analyze and interpret the results of some of
the simulations described by \citet{DMTF}. The mass transfer rates
that these simulations can resolve are much higher than the
Eddington critical rate and probably much higher than the rates
likely to arise during the onset of mass transfer in most realistic
cases. Nevertheless they describe correctly the dynamical aspects of
the mass transfer and tidal interactions under these conditions.
Comparing the predictions of the OAEs with the simulations, we were
pleased to find that they predict the outcome of the simulations
reasonably well.

\acknowledgements This work has been supported in part by NASA's ATP
program grants NAG5-8497 and NAG5-13430.  We thank Joel E. Tohline,
Patrick M. Motl and Ravi Kopparapu for helpful discussions. Finally,
we would like to acknowledge the referee for constructive comments.

\appendix

\section{The Effective Mass-Radius Exponent $\zeta_2$}

\label{Appzeta}

We derive here a simple analytic approximation to the effective
mass-radius exponent when the response of the donor is a combination
of the adiabatic and thermal adjustments to mass loss. Our starting
point is the same as in \S 3 of \citet{DMR}, namely
\begin{equation}
\frac{\d\ln{R_2}}{\d t} = \left(\frac{\partial\ln{R_2}}{\partial
t}\right)_{\rm th}+ \left(\frac{\partial\ln{R_2}}{\partial
t}\right)_{\rm nuc}+\zeta_s\frac{\dot M_2}{M_2}\, , \label{AppStart}
\end{equation}
where the first two terms on the r.h.s. represent the effects of
thermal relaxation and nuclear evolution, and $\zeta_s$ is the
purely adiabatic mass-radius exponent. The thermal relaxation term
may arise as a result of initial conditions, nuclear evolution and
mass transfer, and it is usually not possible to disentangle these
effects if they operate on similar timescales. However, an
approximate description of the radial evolution can be obtained by
viewing it as the result of the superposition of thermal relaxation
of initial conditions and nuclear evolution, plus a thermal
adjustment to mass loss, as follows
\begin{equation}
\frac{\d\ln{R_2}}{\d t} = \nu_2 + \nu'_2+\zeta_s\frac{\dot
M_2}{M_2}\, , \label{Appnus}
\end{equation}
where $\nu_2$ is the superposition of intrinsic thermal and nuclear
evolution, while $\nu'_2$ stands for the rate of thermal radial
reaction to mass loss. Our goal is to combine the last two terms on
the r.h.s. by absorbing the thermal adjustment to mass loss into an
effective mass-radius exponent. We write
\begin{equation}
\zeta_2\frac{\dot M_2}{M_2} = \nu'_2 +\zeta_s \frac{\dot M_2}{M_2}
\label{AppR2dot}
\end{equation}
where $\zeta_2$ is the effective mass-radius exponent we seek. With
the above interpretation, the term $\nu_2$ in eq. (\ref{R2dot})
represents intrinsic thermal and nuclear processes which may operate
on timescales that differ from the standard thermal relaxation rate.
This approach is consistent only if $\nu_2<<\zeta_2(\dot M_2/M_2)$.
Therefore, in what follows, we shall only consider thermal
relaxation in response to mass loss, and set $\nu_2=0$. As a
consequence of the mass loss, the donor's radius $R_2$ will differ
from the equilibrium radius corresponding to its instantaneous mass
$R_{\rm eq}(M_2)$. With these definitions we write
\begin{equation}
\frac{\dot R_2}{R_2} = \frac{R_{\rm eq}(M_2)-R_2}{R_2\tau'} +\zeta_s
\frac{\dot M_2}{M_2} \label{AppR2dot2}
\end{equation}
where we have adopted a simple linear approximation for the thermal
reaction term and $\tau'$ is the corresponding timescale. The
secular evolution of the binary takes place on a mass-transfer
timescale $\tau_{M_2}=-M_2/\dot M_2$. Differentiating  eq.
(\ref{AppR2dot2}) with respect to time, we get
\begin{equation}
\frac{\d^2\ln{R_2}}{\d t^2} =  \frac{1}{\tau'}\frac{R_{\rm eq}}{R_2}
\left(\frac{\zeta_{\rm eq}}{\tau_{M_2}} -
\frac{\zeta_2}{\tau_{M_2}}\right)-\frac{\zeta_s}{\tau_{M_2}^2}
\label{AppR2ddot}
\end{equation}
If the effective mass-radius exponent is to have any meaning, it
must not change much over the evolutionary phase one is considering.
Thus, we require
\begin{equation}
\frac{\d^2\ln{R_2}}{\d t^2} = -\zeta_2/\tau_{M_2}^2 \, .
\end{equation}
Finally, setting $R_{\rm eq}=R_2$ in eq. (\ref{AppR2ddot}), and
solving for $\zeta_2$, we obtain
\begin{equation}
\zeta_2=\frac{\zeta_{\rm eq} +
\zeta_s\tau'/\tau_{M_2}}{1+\tau'/\tau_{M_2}}.
\end{equation}
This expression shows that if the evolution is much slower than the
thermal relaxation ($\tau'\ll\tau_{M_2}$), the donor radius follows
the equilibrium radius closely, whereas if mass transfer occurs
rapidly, the donor reacts adiabatically. The above discussion may be
applied to cataclysmic variables to describe approximately how the
donor becomes increasingly oversized as the orbital period decreases
because $\nu_2\approx 0$. If relaxation from initial conditions or
significant nuclear evolution is taking place on timescales
comparable to either $\tau'$ or $\tau_{M_2}$, the above discussion
is strictly not valid.

\end{document}